\documentclass[%
twocolumn,nofootinbib,amsmath,prd,aps,superscriptaddress,
tightenlines,preprintnumbers,nobalancelastpage]{revtex4-2}

\usepackage{graphicx}
\usepackage{dcolumn}
\usepackage{bm}
\usepackage[utf8]{inputenc}
\usepackage[DIV=13]{typearea}
\usepackage{ragged2e}
\usepackage{calligra,amsmath,amsfonts,bbm,mathrsfs,amssymb,mathtools,amsthm}
\usepackage{slashed,cancel}
\usepackage{hyperref}
\hypersetup{colorlinks=true,urlcolor=blue,anchorcolor=blue,citecolor=blue,filecolor=blue,
            linkcolor=blue,menucolor=blue, linktocpage=true,pdfproducer=medialab}
\usepackage[english]{babel}
\usepackage{indentfirst}
\usepackage{graphicx}
\usepackage{float}
\usepackage{enumerate}
\usepackage{subcaption}
\usepackage{color}
\usepackage{multirow}
\usepackage[normalem]{ulem} 
\usepackage{booktabs}
\usepackage{braket}
\usepackage{MnSymbol}
\usepackage[skins,theorems]{tcolorbox}
\usepackage[capitalise]{cleveref}
\usepackage[compat=1.1.0]{tikz-feynman}
\usepackage{amsmath,amssymb}
\usepackage{orcidlink}
\usepackage[font=small,labelfont=bf,justification=centerlast]{caption}
\usepackage[capitalise]{cleveref}
\newcommand*{\Scale}[2][4]{\scalebox{#1}{$#2$}}%

\tcbset{highlight math style={enhanced,
  colframe=red,colback=white,arc=0pt,boxrule=1pt}}
\numberwithin{equation}{section}

\definecolor{rosso}{cmyk}{0,1,1,0.4}
\definecolor{rossos}{cmyk}{0,1,1,0.55}
\definecolor{rossoc}{cmyk}{0,1,1,0.2}
\definecolor{blu}{cmyk}{1,1,0,0.3}
\definecolor{blus}{cmyk}{1,1,0,0.6}
\definecolor{bluc}{cmyk}{1,1,0,0.1}
\definecolor{verde}{cmyk}{0.92,0,0.59,0.25}
\definecolor{verdec}{cmyk}{0.92,0,0.59,0.15}
\definecolor{verdes}{cmyk}{0.92,0,0.59,0.4}
\makeatletter
\def\@ssect@ltx#1#2#3#4#5#6[#7]#8{%
  \def\H@svsec{\phantomsection}%
  \@tempskipa #5\relax
  \@ifdim{\@tempskipa>\z@}{%
    \begingroup
      \interlinepenalty \@M
      #6{%
       \@ifundefined{@hangfroms@#1}{\@hang@froms}{\csname @hangfroms@#1\endcsname}%
       {\hskip#3\relax\H@svsec}{#8}%
      }%
      \@@par
    \endgroup
    \@ifundefined{#1smark}{\@gobble}{\csname #1smark\endcsname}{#7}%
  }{%
    \def\@svsechd{%
      #6{%
       \@ifundefined{@runin@tos@#1}{\@runin@tos}{\csname @runin@tos@#1\endcsname}%
       {\hskip#3\relax\H@svsec}{#8}%
      }%
      \@ifundefined{#1smark}{\@gobble}{\csname #1smark\endcsname}{#7}%
      \addcontentsline{toc}{#1}{\protect\numberline{}#8}%
    }%
  }%
  \@xsect{#5}%
}%

\newcommand\subsetsim{\mathrel{%
  \ooalign{\raise0.2ex\hbox{$\subset$}\cr\hidewidth\raise-0.8ex\hbox{\scalebox{0.9}{$\sim$}}\hidewidth\cr}}}
\newcommand*{\diff}[1]{\text{d}#1}

\let\vec\mathbf

\graphicspath{{Draft_Plots/}}

\begin{document}
\preprint{IPPP/24/75}
\preprint{ CERN-TH-2024-205}

\title{\Large Extra-dimensional axion patterns}

\author{Arturo de Giorgi \orcidlink{0000-0002-9260-5466
}}
 \email{arturo.de-giorgi@durham.ac.uk}
\affiliation{Institute for Particle Physics Phenomenology, Durham University, South Road, DH1 3LE, Durham, UK
}

\author{Maria Ramos \orcidlink{0000-0001-7743-7364}}
\email{maria.ramos@cern.ch}
\affiliation{CERN, Theoretical Physics Department, Esplanade des Particules 1, Geneva 1211, Switzerland}

\begin{abstract}
We study the~\textit{complete} parameter space of a bulk axion in flat and warped extra spacetime dimensions. We characterize in detail the regimes where no single KK mode is produced along the canonical QCD axion line, and instead, it is maximally deviated along with several other axions that constitute a multiple solution to the strong CP problem.
In both flat and Randall-Sundrum scenarios, and assuming that all Peccei-Quinn breaking comes from QCD, we find that these solutions are however subject to tight phenomenological constraints. 
In light of these results, we expect that only KK canonical patterns (with the zero-mode close to the standard QCD line) can emerge from a bulk axion in one or more extra spacetime dimensions.
As a byproduct, we generalize the axions eigenvalue and eigenvector equations for an arbitrary number of spacetime dimensions and compactifications.
\end{abstract}

\maketitle

\tableofcontents
\newpage
\section{Motivation and summary}
Axions are among the most compelling candidates for new physics, and today, one of the most prominent targets of small-scale experiments built all over the world. They are associated with continuous shift-symmetries, which can be broken by non-perturbative dynamics, endowing the axions with a mass $m_{i}^2 \sim \Lambda_{\rm ins}^4 /f_i^2$, where $\Lambda_{\rm ins}$ denotes an instanton scale. Therefore, they can be naturally light and weakly interacting while providing a window to deep UV physics, that might otherwise be beyond our reach. This connection is further strengthened by quantum gravity consistency requirements and string theory compactifications~\cite{WITTEN1984351,Arvanitaki:2009fg,Heidenreich:2020pkc}, that lead us to expect a plethora of axions at low energy populating diverse scales.
Following the Peccei-Quinn (PQ) paradigm, the strong CP problem could be solved if one combination of these fields remains light at the QCD scale, to dynamically explain the absence of CP violation in the strong sector.
This motivates searches for the QCD axion in a precise band where $m_a^2 f_a^2 = \Lambda_{\rm QCD}^4$.

\begin{figure*}[]
    \centering
    \includegraphics[width=0.8\textwidth]{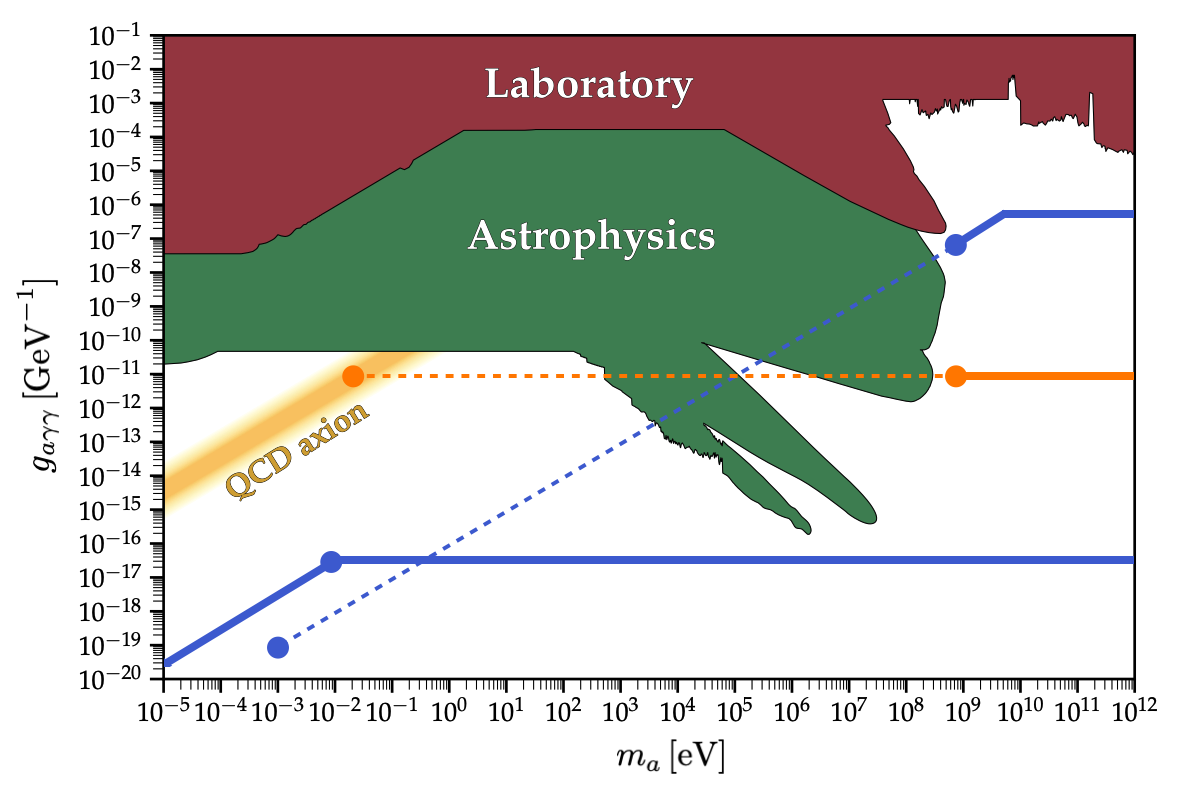}
    \caption{\it Current axion bounds~\cite{AxionLimits} and schematic representation of possible KK axion patterns arising from extra-dimensions. 
    The dashed lines highlight gaps in the mass spectrum.}
    \label{fig:schematic}
\end{figure*}
Recently, it has been shown that if such light combination mixes with other fields, the solutions to the strong CP problem could be located outside the canonical single axion band, with the maximal departure possible given by $g_i \equiv m_i^2 f_i^2 /\Lambda_{\rm QCD}^4 = n_\star$, the number of QCD axions in Nature~\cite{Gavela:2023tzu}. Following the previous work, we denote the axions satisfying this mass-scale relation by QCD \textit{maxions}. Such a scenario leads to a new displaced QCD axion band where all the $n_\star$ signals are aligned, which is a direct consequence of the QCD axion sum rule~\cite{Gavela:2023tzu}:
\begin{equation}
    \sum_{i=1}^{n_\star} \frac{1}{g_i} = 1\,,
    \label{eq:sum-rule}
\end{equation}
that has been derived for generic mixing potentials that preserve a PQ symmetry at the classical level.
Finding a compelling UV framework where such large deviations are realized could radically change the axion phenomenology by redefining the target of many axion experiments, without the need to extend the SM gauge group. However, the first difficulty that one encounters in trying to build such a setup is that the large deviations must then be associated with a large number of axions that mix sizably into each other. While this is challenging from a 4D perspective, such a large number of axions is a direct prediction in models with extra spacetime dimensions, where one axion propagating in the bulk of the extra dimension is identified with an infinite tower of Kaluza-Klein (KK) modes upon dimensional reduction~\cite{Kaluza:1921tu,Klein:1926tv}. We are therefore led to search for the non-canonical axions in such scenarios, assuming that the SM is localized on a 4D brane.

Phenomenological studies of a 5D QCD axion have been presented in a series of seminal works~\cite{Dienes:1999gw,DiLella:2000dn,Flacke:2006ad} and have garnered renewed attention in recent years~\cite{ Anastasopoulos:2018uyu,Cox:2019rro, Gherghetta:2020keg, Bonnefoy:2020llz, Gendler:2024gdo, Agrawal:2024ejr,Craig:2024dnl}. One motivation for this framework is to produce a weak axion coupling to gauge bosons from a high fundamental PQ scale, by the same mechanism that lowers the Planck scale in the higher dimensional theory. Importantly, this framework also allows to decouple the mass of the zero mode from the PQ scale, $m_{\rm PQ}$, which can be instead set by the scale of the extra dimension, $R^{-1}$, as first pointed out in Ref.~\cite{Dienes:1999gw}. Moreover, the presence of the KK tower can impact substantially the experimental bounds, as well as provide a mechanism for dynamical dark matter in which the relic abundance is shared among several of the KK fields. These aspects have been discussed extensively in the literature in the context of flat extra-dimensional models~\cite{Dienes:2011ja, Dienes:2011sa,Dienes:2012jb}.

In this work, we will focus on the critical role of the KK tower in solving the strong CP problem in the full parameter space of both flat and curved Randall-Sundrum (RS)~\cite{Randall:1999ee,Randall:1999vf} scenarios.
We will prove that an extra-dimensional bulk axion can either give rise to $1$ canonical axion or a set of $N$ deviated QCD maxions predicted by the sum rule~\eqref{eq:sum-rule}, that is~\textit{maxions are the non-canonical solutions of generic extra spacetime dimensional models} (assuming that QCD is the only source of PQ breaking). 
The two possible patterns that can arise are shown schematically in Fig.~\ref{fig:schematic}, represented by an orange or a blue line. The first indicates a canonical QCD axion solution, where all but one mode are decoupled from the solution to the strong CP problem. The second type of pattern corresponds to the extra-dimensional QCD maxion scenario, 
where no single axion is found close to the canonical band. 
Such maximal mixing regimes arise if $A \,\mu_1/m_{\rm PQ} \lesssim 1$, with $\mu_1$ denoting the lightest graviton mass and $A$ the warping factor associated with the curvature of the extra dimension. Such curvature can take the QCD maxions further out of the canonical line, and separate significantly the zero-mode from the rest of the KK tower; this is representated by the dashed blue line in Fig.~\ref{fig:schematic}. 

The aim of our work is to, first, characterize fully the axion patterns stemming from extra dimensions and show how the location of the KK modes encodes a mapping to the fundamental properties of the higher-dimensional bulk axion. In flat extra-dimensional scenarios, the mass spectrum and the KK couplings have been previously computed in a series of works~\cite{Dienes:1999gw,Dienes:2011ja,Dienes:2011sa}. Nonetheless, we will obtain a general formalism from which we will re-derive these results and extend them to the case of a warped extra dimension. Moreover, we will obtain generalized expressions for the eigenvalues and eigenvector equations that apply to a broader class of
higher-dimensional axion models with arbitrary compactifications and number of spacetime dimensions, whose relevance extends beyond the scope of our work.

Secondly, we will take into account the most constraining phenomenological probes, to identify which of these patterns are allowed by current data. We will be mostly interested in understanding whether the extra-dimensional maxion scenarios, with $g_i \gg 1$, survive this scrutiny. Such scenarios are the most interesting ones in what concerns the possibility of identifying new regions of signal to search for QCD axions. Moreover, some of the patterns represented in Fig.~\ref{fig:schematic} are common to other frameworks, namely the string axiverse~\cite{Gendler:2023kjt} and Grand Unified Theories~\cite{Agrawal:2022lsp}. (This is not surprising, as these patterns follow directly from the generic QCD axion sum rule.) For this reason, investigating the constraints on KK signals from extra-dimensional models is essential to eventually discriminate the origin of a potential multiple axion pattern observed in Nature.

We will show that the QCD maxion patterns from a PQ bulk field are tightly constrained by a combination of consistency, astrophysical and gravitational bounds. To the best of our knowledge, this result is completely new for scenarios with a warped extra dimension. Moreover, this result holds independently of the number of (universal) spacetime dimensions and the VEV profile of the PQ field in the bulk.
We, therefore, expect that more complex constructions, with additional bulk fields or less generic compactifications, are required to produce the most exotic patterns in Fig.~\ref{fig:schematic}. In turn, the observation of a plateau of heavy fields together with a canonical QCD axion could be explained by a single PQ field, propagating in the bulk of a hidden dimension. We will demonstrate how a potential observation of such a plateau could in this case 
hint towards the location of the zero mode in the canonical QCD axion band.

\section{Extra dimensional models}
We consider the Einstein-Hilbert action in 
${d=4+\delta}$ spacetime dimensions
    \begin{equation}
    \label{eq:RS_Action}
            S_{\text{EH}} = \frac{M_d^{2+\delta}}{2}\int \diff{}^4x \int\limits_{-\pi R}^{\pi R} \diff{y_1}\,\dots\int\limits_{-\pi R}^{\pi R} \diff{y_\delta} \sqrt{g}(\mathcal{R}-2\Lambda_B ) 
            \,,
    \end{equation}
with $y_a$ denoting a coordinate of the extra dimensions, that we assume to be compactified on a $(S^1/\mathbb{Z}^2)$ orbifold of universal radius $R$. Accordingly, $y_a\to -y_a$ are identified for all points in the interval $[-\pi R, \pi R]$.
In the equation above, $g$ is the determinant of the metric, $\mathcal{R}$ the Ricci scalar, $M_d$ the $d$-dimensional Planck mass, and $\Lambda_B$ the vacuum energy of the bulk~\footnote{We will denote coordinates belonging to the full extra-dimensional models, to the 4D EFT and to the extra-dimensions with capital indices (e.g. $x^A$), greek letters (e.g. $x^\mu$ with $\mu=0,1,2,3$) and in latin lower case (e.g. $x^a$  with $a=5,6,\dots,\,d$, or $y_a$ with $a=1,\dots\,,\delta$), respectively. We will also make use of $\vec{y}=(y_1,\dots\,,y_\delta)$ to highlight the full set of extra-dimensional coordinates.}. Regarding the matter content, we focus on models where the SM fields are localized on the branes, that is the boundaries of the extra dimensions.

The action in Eq.~\eqref{eq:RS_Action} features an extra-dimensional graviton, $h_{MN}(x,\vec{y})$.
As usual, to obtain the effective theory in four spacetime dimensions, the $\delta$-dimensional fields can be decomposed into a tower of KK modes via a
set of orthonormal functions $\{\psi_{\Vec{n}}(\vec{y})\}$; in the case of the graviton, 
\begin{equation}
    h_{\mu\nu}(x,\vec{y})=\dfrac{1}{(2\pi R)^{\delta/2}}\sum\limits_{\vec{n}}^\infty h^{(\vec{n})}_{\mu\nu}(x)\psi_{\vec{n}}(\vec{y})\,.
\end{equation}
The relation between the extra-dimensional and 4D Planck mass can then be obtained by matching the 4D EFT to General Relativity~(GR), and ultimately amounts to a rescaling controlled by the volume of the extra dimensions ($V_\delta$):
\begin{equation}
\label{eq:planck-mass-def}
    m_{P}^2=V_\delta\,M_d^{2+\delta}=\dfrac{(2\pi R)^\delta}{\psi_{\vec{0}}^2}\,M_d^{2+\delta}\,,
\end{equation}
where $m_P$ is the reduced Planck mass.

On top of the gravitational interactions, we will introduce an axion field that propagates in the bulk of all dimensions. (We  discuss later departures from this assumption.) The corresponding action for such a field is:
\begin{align}
        S_a & = \int \diff{}^4x \int\limits_{-\pi R}^{\pi R} \diff{y_1}\,\dots\int\limits_{-\pi R}^{\pi R} \diff{y_\delta} \sqrt{g}\left[\dfrac{1}{2}M_s^\delta g^{AB}\partial_A a \partial_B a \nonumber \right. \\
       & \left. +\dfrac{\alpha_s}{8\pi}\dfrac{a}{f_d}G_{\mu\nu}\widetilde{G}^{\mu\nu}\delta^{(\delta)}(\Vec{y}-\pi R)\right]\,,
        \label{Eq:ActionAxiondDim}
    \end{align}
    where the $\delta$-function localizes the axion couplings to QCD on the SM brane, at $y_a = \pi R$. 
    The interactions above are assumed to be generated at a fundamental scale $M_s$, with $f_d$ denoting the PQ breaking scale. The former can be removed from the kinetic term by an appropriate redefinition of the axion field and the axion decay constant.
    
    Analogously to the graviton case, the extra-dimensional axion can be decomposed as
    \begin{equation}
        a(x,\vec{y})=\dfrac{1}{(2\pi R M_s)^{\delta/2}}\sum\limits_{\vec{n}} \hat a_{\vec{n}}(x)\psi_{\Vec{n}}(\vec{y})\,.
        \label{eq:decomposition}
    \end{equation}
The canonical kinetic and mass terms for these KK states are obtained by imposing the following normalization conditions:
    \begin{align}
    \label{eq:norm-conds}
        &\dfrac{1}{(2\pi R)^{\delta}}\int\limits_{-\pi R}^{\pi R} \diff{\vec{y}}\sqrt{g}g^{\mu\nu}\, \psi_{\Vec{n}}(\vec{y})\psi_{\Vec{m}}(\vec{y})=\delta_{\vec{n},\vec{m}}\eta^{\mu\nu}\,,\\
        &\dfrac{1}{(2\pi R)^{\delta}}\int\limits_{-\pi R}^{\pi R} \diff{\vec{y}}\sqrt{g}\, g^{ab}\partial_a\psi_{\Vec{n}}(\vec{y})\partial_b\psi_{\Vec{m}}(\vec{y})=-\mu_{\vec{n}}^2\delta_{\vec{n},\vec{m}}\,,
    \end{align}
where $\mu_{\vec{n}}$ is the mass of the $\vec{n}$-mode obtained by solving the Sturm-Liouville problem of the equations of motions~(EOM) of the free theory
    \begin{equation}
    \label{eq:eom}
        \partial_B(\sqrt{g}g^{AB}\partial_A a)=0\,,
    \end{equation}
that depends on the background metric of the extra dimensions.

    From here on, we restrict the discussion to the case $\delta=1$ (once again, departures from this assumption will be discussed in the next sections of our work). The effective action for the KK modes reads, in this case:
\begin{align}
        S_4&\supset \int \diff{}^4x \sum\limits_n\left(\dfrac{1}{2}(\partial_\mu \hat a_n)^2- \dfrac{1}{2}\mu_n^2 \hat a_n^2\right)\nonumber \\
        &+\dfrac{1}{f_{\rm PQ}}\dfrac{\alpha_s}{8\pi}\left[\sum\limits_n\hat a_n \psi_n(\pi R) \right]\Tilde{G}_{\mu\nu}G^{\mu\nu}\,,
        \label{eq:KKL}
\end{align}
where $f_{\rm PQ}\equiv f_4 \equiv \sqrt{2 \pi R M_s} f_5$ is the effective PQ scale in 4D.
Below, we match the parameters of this EFT to the flat and RS models. The following sections are dedicated to a detailed study of the emergent axion phenomenology.

\subsection{Flat extra dimension}

In a flat extra-dimensional model, the background metric is given by~\footnote{We adopt the ``mostly minus'' convention for the metric.}
       \begin{equation}
        \diff{s}^2=\eta_{\mu\nu}\diff{x}^\mu\diff{x}^\nu-\diff{y}^2\,,
        \end{equation}
    where $\eta_{\mu\nu}$ is the Minkowski flat metric and ${\Lambda_B = 0}$. This leads to the following identification of the reduced Planck mass:
    \begin{equation}
        {m}_{P}^2\equiv (2\pi R) M_5^{3}\,,
        \label{eq:MLED}
    \end{equation}
    such that the theory is effectively ruled by a single parameter.

    The EOM of the 5D axion reads:
    \begin{equation}
        \partial_\mu \partial^\mu a +\partial_5 \partial^5a=0\,.
    \end{equation}
    This reproduces the canonical KG equation if and only if one imposes
    \begin{equation}
        \partial_5 \partial^5\psi_n=\mu_n^2 \psi_n\,.
    \end{equation}
    The solutions to this equation which are compatible with the orbifold symmetry read
    \begin{equation}
        \psi_n(y)=N_n \cos(\mu_ny)\,,
    \end{equation}
    where $N_n$ is fixed by the normalization conditions in Eq.~\eqref{eq:norm-conds}:
     \begin{equation}
        N_n =\begin{cases}
            1 & n=0\,,\\
            (-1)^n\sqrt{2}& n>0\,.
        \end{cases}
    \end{equation}
    Note that a minus sign was included in these factors, for convenience in subsequent steps.
    By imposing the boundary condition $\partial_5 \psi_n(\pi R)=0$, one obtains the mass spectrum of the theory:
    \begin{equation}
    \label{eq:mass-LED}
        \mu_n = \dfrac{n}{R}\,.
    \end{equation}
The WFs on the IR brane, therefore, read:
    \begin{equation}
    \label{eq:IR-LED}
        \psi_n(\pi R)=\begin{cases}
            1 & n=0\,,\\
            \sqrt{2}&n>0\,.
        \end{cases}
    \end{equation}

\subsection{Randall-Sundrum}

In the RS model, the curvature is manifested via a conformal factor in front of the 4D metric:
    \begin{equation}
        \diff{s}^2=e^{-2k|y|}\eta_{\mu\nu}\diff{x}^\mu\diff{x}^\nu-\diff{y}^2\,,
        \end{equation}
        with $k\equiv \sqrt{-\Lambda_B/6}$.
        The exponential $A(y)\equiv e^{-k|y|}$ is typically called the \textit{warp factor}, which reduces to $1$ in the flat case (where $k \to 0$).
    The connection to GR is now encoded by
    \begin{equation}
        {m}_{P}^2\equiv \dfrac{M_5^3}{k}\left(1-e^{-2\mu\pi}\right)\,,
    \end{equation}
    where $\mu\equiv k R$. Therefore, the model is ruled by two parameters. We recall that in flat models, current limits on $R$ are incompatible with a solution to the Higgs hierarchy problem with $\delta=1$. On the contrary, due to the warp factor in this setup, the Higgs mass is expected to be $\sim e^{-\mu\pi} M_5$ in the IR brane, so that the hierarchy problem is solved even for $M_5 \sim m_P$ and $\mu\sim\mathcal{O}(10)$. We will not attempt to solve this problem in the current analysis.

    The RS metric can be expressed using an alternative coordinate, $\diff{y}=A(y)\,\diff{z}$, such that it becomes conformally flat~\footnote{{The relation between the two sets of coordinates can be obtained by integrating $A(y)^{-1}\,\diff{y}=\diff{z}$. One can set $A=e^{-k y}=1/(zk)$ by appropriately choosing the integration boundaries.}}:
    \begin{equation}
        \label{eq:conformal}
        \diff{s}^2=A^2(z)\left(\eta_{\mu\nu}\diff{x}^\mu\diff{x}^\nu-\diff{z}^2\right)\,.
    \end{equation}
    In terms of this coordinate, 
    the EOM for the axion (cfr. Eq.~\eqref{eq:eom}) reads
    \begin{equation}
        \begin{split}
            A^3\partial_\mu\partial^\mu a-\partial_z\left(A^3\partial_z\right)a = 0\,,
        \end{split}
    \end{equation}
    which -- upon KK reduction -- leads to the following eigenvalue equation:
    \begin{equation}
        \partial_z\left(A^3\partial_z\right)\psi_n=-A^3 \mu_n^2 \psi_n\,.
    \end{equation}
    The solutions are given by
    \begin{equation}
        \psi_n(y)=N_n e^{2k|y|}\left[J_2(z_n)+\alpha_n Y_2(z_n)\right]\,,
    \end{equation}
    where
    \begin{equation}
        z_n\equiv \mu_n\dfrac{e^{k|y|}}{k}\,,
    \end{equation}
    and $N_n$ is a normalization constant. The constants $\alpha_n$ and $\mu_n$ are determined by the boundary conditions $\left.\partial_y \psi_n\right|_{y=0,\pi R}=0$. We obtain:
    \begin{equation}
        \alpha_n=\dfrac{J_1\left(\mu_n\dfrac{e^{\mu\pi}}{k}\right)}{Y_1\left(\mu_n\dfrac{e^{\mu\pi}}{k}\right)}\,,
    \end{equation}
    while the masses $\mu_n$ are the solutions of the transcendental equation
    \begin{equation}
        J_1\left(\mu_n\dfrac{e^{\mu\pi}}{k}\right)Y_2\left(\dfrac{\mu_n}{k}\right)-J_2\left(\dfrac{\mu_n}{k}\right)Y_1\left(\mu_n\dfrac{e^{\mu\pi}}{k}\right)=0\,.
    \end{equation}
    There is no general solution to this equation,
    but in the limit $e^{\mu\pi}\gg 1$ we find:
    \begin{align}
        \psi_n(\varphi = y R^{-1})& \approx -\sqrt{2\pi \mu}\dfrac{e^{\mu(2|\varphi|-\pi)}}{J_0(\gamma_n)}J_2(\gamma_n e^{\mu(|\varphi|-\pi)})\,, \nonumber\\
        \mu_n & \approx \gamma_n k e^{-\mu\pi}\label{eq:mass-RS}\,,
    \end{align}
    where $\gamma_n\approx \pi(n+1/4)$ is the $n$th zero of the Bessel function $J_1(x)$. In the opposite limit, i.e. $k\ll 1$, the setup reduces to the flat model.
    On the IR brane, these WFs read:
    \begin{equation}
    \label{eq:IR-RS}
       \psi_n(\pi R)\simeq \sqrt{2\pi\mu} \times \begin{cases}
           1& n=0 \,,\\
           e^{\pi\mu} & n>0\,,
       \end{cases}
    \end{equation}
    where we have used the identity $J_2(\gamma_n)=-J_0(\gamma_n)$.

\section{General mass matrix}
The EFT in Eq.~\eqref{eq:KKL} predicts a very particular mixing structure for the KK axions, that in full generality can be written as:
\begin{equation}
\label{eq:general-mm}
    \begin{split}
        \left(\mathbf{M}^2\right)_{ij}&=m_\text{PQ}^2\left[\psi_i \psi_j +\mathsf{y}^2\left(\dfrac{\mu_i}{\mu_1}\right)^2\delta_{ij}\right]\,,
        \end{split} 
\end{equation}
i.e.
\begin{equation*}
    \frac{\mathbf M^2}{m_\text{PQ}^2} 
        =\begin{pmatrix}
        \psi_0^2 & \psi_0 \psi_1  & \psi_0 \psi_2  &\dots\quad\\
        \psi_1 \psi_0  & \psi_1^2+\mathsf{y}^2 & \psi_1 \psi_2 &\dots\quad\\
        \psi_2 \psi_0  & \psi_2\psi_1  & \psi_2^2 +\mathsf{y}^2\left(\dfrac{\mu_2}{\mu_1}\right)^2 &\dots\quad\\ 
        \dots & \dots & \dots & \dots\quad
    \end{pmatrix}\,,
\end{equation*}
where
\begin{align}
    & \mathsf{y}\equiv\dfrac{\mu_1}{m_\text{PQ}}\,, &&\psi_i\equiv \psi_i(\pi R)\,,
\end{align}
$m_{\rm PQ}^2 \equiv \chi_{\rm QCD}/f_{\rm PQ}^2$ is the standard PQ mass and 
\begin{equation}
\chi_\text{QCD} \equiv m_\pi^2 f_\pi^2 \frac{m_u m_d}{\left(m_u + m_d\right)^2}
\end{equation}
is the QCD topological susceptibility. Such mass matrix reproduces any of the models described in the previous section, by simply matching the WFs to Eqs.~\eqref{eq:IR-LED} (flat) and~\eqref{eq:IR-RS} (RS).
Note that the main difference between these constructions in the $\mu\gg 1$ limit lies not in the mass ratios, but rather in the expressions for these WFs, which are exponentially enhanced by the curvature.

We aim to diagonalize this mass matrix exactly. Even though this is challenging in general, the problem can be simplified under the assumption that
\begin{equation}
   \psi_i=\psi_j\,,\qquad \forall\, i,j>0\,,
\end{equation}
which holds in both the flat and RS models in the interesting limits. 
Under this assumption, we find that the eigenvalues $m_\lambda^2 \equiv \lambda^2 m_{\rm PQ}^2 $ are the solutions to the following equation:
\begin{equation}
\label{eq:eigenvalue-eq}
        {\sum\limits_{n=0}^\infty\dfrac{(\psi_n)^2}{\lambda^2-(\mu_n/\mu_1)^2\mathsf{y}^2} =1\,;}
\end{equation}
see App.~\ref{app:eigenvalues} for details. We note that while this formula has been derived in Ref.~\cite{Dienes:1999gw} for the flat scenario, here we prove that it holds also for RS models.
A more general approach to obtain the eigenvalues will also be derived in Sec.~\ref{sec:generalizations}.

\section{Maximal displacements of KK axions}

To identify all possible patterns of KK axions in the mass vs. coupling parameter space, we turn to the computation of the $g_i$ factors, as defined in Ref.~\cite{Gavela:2023tzu}. 

Let us denote by $U$ the rotation matrix that diagonalises the mass matrix in Eq.~\eqref{eq:general-mm}, such that the physical fields are identified via the relation 
\begin{equation}
    \hat{a}_i=\sum\limits_\lambda U_{i\lambda}a_\lambda\,.
\end{equation}
The physical couplings in the mass basis,
\begin{equation}
    \mathcal{L}\supset\dfrac{\alpha_s}{8\pi}\dfrac{a_\lambda}{f_\lambda}G\Tilde{G}\,,
\end{equation}
are given by
\begin{equation}
    \dfrac{1}{f_\lambda}=\dfrac{1}{f_\text{PQ}} \left(\sum\limits_{n=0}^\infty \psi_n\ U_{n \lambda}\right) \,.
\end{equation}
Consequently, the $g$-factor of each eigenstate reads
\begin{equation}
\label{eq:g-factors-computation}
    g_\lambda \equiv \frac{m_\lambda^2 f_\lambda^2}{\chi_{\rm QCD}} = \left(\dfrac{\lambda}{\sum\limits_{n=0}^\infty  \psi_n \, U_{n \lambda}}\right)^2\,.
\end{equation}

The detailed calculation of the rotation matrix is presented in App.~\ref{app:eigenvectors}. Using the results therein, we find:
\begin{align}
    &U_{i\lambda}=\mathcal{N}_\lambda\,\dfrac{\psi_i}{\lambda^2-(\mu_i/\mu_1)^2\mathsf{y}^2}\,,
\end{align}
    with
        \begin{align}
   {\mathcal{N}_\lambda=\left[\sum\limits_{n=0}^\infty\left(\dfrac{\psi_n}{\lambda^2-(\mu_n/\mu_1)^2\mathsf{y}^2}\right)^2\right]^{-1/2}\,.}
   \label{eq:Nlambda}
\end{align}

The $g$-factors can now be found using Eqs~\eqref{eq:g-factors-computation} and \eqref{eq:eigenvalue-eq}:
\begin{align}
        g_\lambda & =\dfrac{\lambda^2}{\mathcal{N}_\lambda^2}=\sum\limits_{n=0}^\infty\left(\dfrac{\lambda\, \psi_n}{\lambda^2-(\mu_n/\mu_1)^2\mathsf{y}^2}\right)^2\,.
        \label{eq:gfactor}
\end{align}
Remarkably, using the definition above, one rediscovers the QCD axion sum rule in the extra-dimensional models:
\begin{equation}
    \sum\limits_\lambda \left(\dfrac{1}{g_\lambda}\right)= \sum\limits_\lambda \left(\dfrac{\mathcal{N}_\lambda^2}{\lambda^2}\right)=1\,,
    \label{Eq:SumRule}
\end{equation}
proven in all generality in Ref.~\cite{Gavela:2023tzu}.  
In the flat scenario, this result has been previously derived in Ref.~\cite{Dienes:1999gw}. In App.~\ref{app:differential-g-factors} we present a different proof, that extends beyond the flat case.
We will now infer the consequences of these general results to the models introduced in the previous section.

\subsection{QCD maxions in flat space}
In {flat scenarios}, the sum in Eq.~\eqref{eq:eigenvalue-eq} can be performed exactly:
\begin{equation}
    \dfrac{\pi \lambda}{\mathsf{y}}\cot\left(\dfrac{\pi \lambda}{\mathsf{y}}\right)=1+2\left(\dfrac{\lambda^2-\psi_0^2}{\psi_1^2}\right)\,.
\end{equation}
By inserting the values of $\psi_{0,1}$, we obtain
\begin{equation}
    \dfrac{\pi \lambda}{\mathsf{y}}\cot\left(\dfrac{\pi \lambda}{\mathsf{y}}\right)=\lambda^2\,,
    \label{eq:lambdas}
\end{equation}
in agreement with what was previously found in Ref.~\cite{Dienes:1999gw}. 
Similarly, one finds: 
\begin{align}
    g_\lambda^{\text{flat}}& =\frac{1}{2} \left(\frac{\pi ^2 \psi_1^2}{2\mathsf{y}^2}+\frac{3\lambda^2- \psi_0^2}{\lambda ^2}+\frac{2 \left(\lambda ^2-\psi_0^2\right)^2}{\lambda ^2 \psi_1^2}\right)\,\nonumber\\
& = \dfrac{1}{2}\left(\lambda^2+1+(\pi/\mathsf{y})^2\right)\,.
    \label{eq:GLED}
\end{align}
This formula encodes some of possible patterns of axions represented in Fig.~\ref{fig:schematic}.
To see how, let us first consider the limit $\mathsf y\gg 1$. In this case, the expression above reduces to $\lambda_0 \sim 1$, such that $g_0 = 1$. The zero mode therefore behaves as the canonical QCD axion, being e.g. localized in the same position as the orange bullet in Fig.~\ref{fig:schematic}.

On the contrary, maxion patterns can appear only when the mass matrix is not diagonal, i.e. when the diagonal elements are subleading with respect to the off-diagonal ones, for a subset of modes $n< n_\star$. This leads to the condition:
\begin{equation}
\label{eq:LED-regime}
    \mathsf{y} \ll \dfrac{\mu_1}{\mu_{n_\star}}\psi_1 \,,
\end{equation}
which, in the present model, requires $\mathsf y \lesssim 1$. 
In such limit, we find the masses to be well-described by
\begin{equation}
    \lambda_n = \left(n+\dfrac{1}{2}\right) \mathsf{y}\left[1-\dfrac{\mathsf{y}^2}{\pi^2}+\mathcal{O}(\mathsf{y}^4)\right]\,.
    \label{eq:lamapprox}
\end{equation}
Hence, for the zero mode, we have:
\begin{align}
    &\lambda_0 \approx \frac{\mathsf{y}}{2}\,,\\
    &g_0\approx \frac{\pi ^2}{2 \mathsf{y}^2}\,,\\
    &
    f_0  \approx f_4\times\,\sqrt{2}\dfrac{\pi^2}{\mathsf{y}^2}\,.
\end{align}
Indeed, in this limit, the third term in Eq.~\eqref{eq:GLED} dominates over the mass term: this is not only true for the zero mode but for several of the KK axions in the tower.
We, therefore, expect to find $n_{\star} \sim g_0$ QCD maxions with the same $g$-factor and deviated from the canonical line by a factor of $\sqrt{g_0}$.
Technically, these are approximately maxions as their $g$-factors are almost identical, but not exactly equal. Such behaviour can be checked explicitly by using the same Eq.~\eqref{eq:lamapprox} for the first $n$-eigenvalues 
as long as the condition~\eqref{eq:LED-regime} is satisfied; we find:
\begin{align}
    &\lambda_{n>0}\approx\left(n+\dfrac{1}{2}\right) \mathsf{y}\,,\\
    &g_{n>0}\approx \frac{\pi ^2}{2 \mathsf{y}^2}\,,\\
    &f_{n>0}\approx  f_4\times\,\dfrac{\sqrt{2}}{1+2n}\dfrac{\pi^2}{\mathsf{y}^2} \,.
\end{align}
These results confirm our expectations and agree with previous studies of the mass spectrum of this theory~\cite{Dienes:2011ja}.

The tower of QCD maxions is always accompanied by a plateau of heavier modes that decouple from the sum rule. Since the QCD contribution to the mass is negligible for these modes, i.e. they are already eigenstates in the basis of Eq.~\eqref{eq:KKL}, we must find $\lambda_\text{plateau} \sim n \mathsf{y}$. 
The plateau is then expected to appear for modes with $n\gtrsim n_\star \approx \pi/\mathsf{y}^2$, for which the $g$-factors are dominated by the mass term in Eq.~\eqref{eq:GLED}. (Note that such modes no longer satisfy the condition for large mixings; see Eq.~\eqref{eq:LED-regime}.) Working out the expressions explicitly, we obtain:
\begin{align}
    &\lambda_{\rm plateau}\approx n \mathsf y\,,
    \label{eq:mnLED}\\
    &g_{\rm plateau}\approx\dfrac{n^2 \mathsf y^2}{2}=\dfrac{n^2 \mathsf y^2}{\psi_1^2}\,,\\
    &f_{\rm plateau}\approx  \frac{f_4}{\sqrt 2}=\dfrac{f_4}{\psi_1}\,.
    \label{eq:fplateauLED}
\end{align}
These are consistent with the previous arguments and show
that the plateau's contribution to the sum rule, $\sum_{n=n_\star}^\infty 1/g_n \sim \sum_{n=n_\star}^\infty 1/(n \mathsf y)^2$, falls rapidly to zero {ensuring its convergence}.

By performing a numerical analysis, we have obtained representations of the eigenmodes for different values of $\mathsf{y}$, as shown in Fig.~\ref{fig:LED}. While the observation of $n_\star$ aligned axions would be a probe of the extra dimension scale in units of the Peccei Quinn mass, the observation of the plateau would allow us to extract the coupling $\psi_1/f_4$. Combining these two features, we would be able to determine the value of $R$ corresponding to the pattern of axions observed.
This scale is determinant to understand whether such a pattern complies with current bounds. While large $R$ values displace the QCD axion canonical band to the right, $f_4$ can enhance the plateau coupling making it visible to experiments.

\subsection{QCD maxions in curved space}
\begin{figure*}[t]
    \centering
    \begin{subfigure}[t]{0.45\textwidth}
    \centering
    \includegraphics[width=\textwidth]{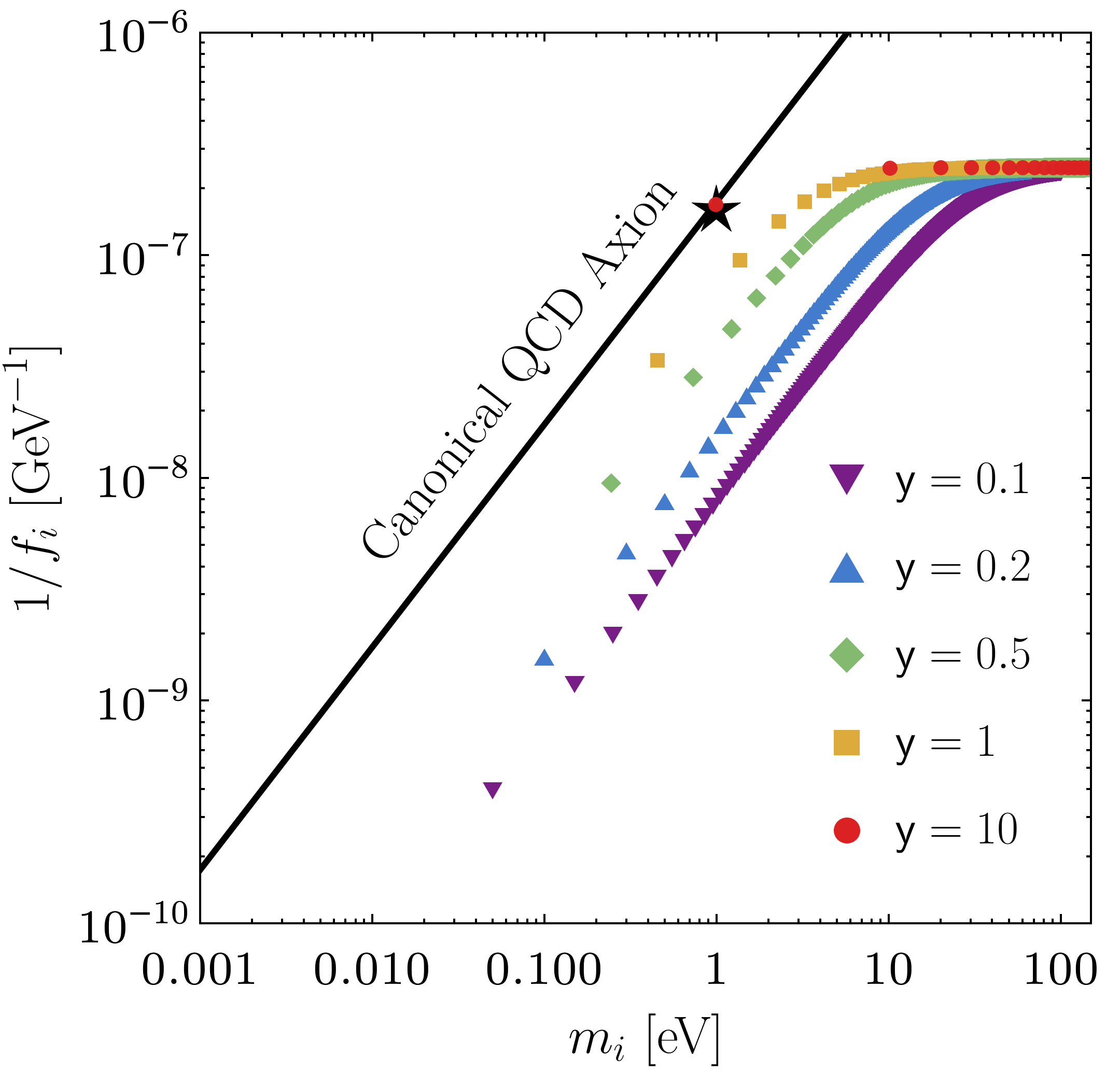}
    \caption{Flat.}
    \label{fig:LED}
    \end{subfigure}
    \hfill
    \begin{subfigure}[t]{0.45\textwidth}
    \centering
    \includegraphics[width=\textwidth]{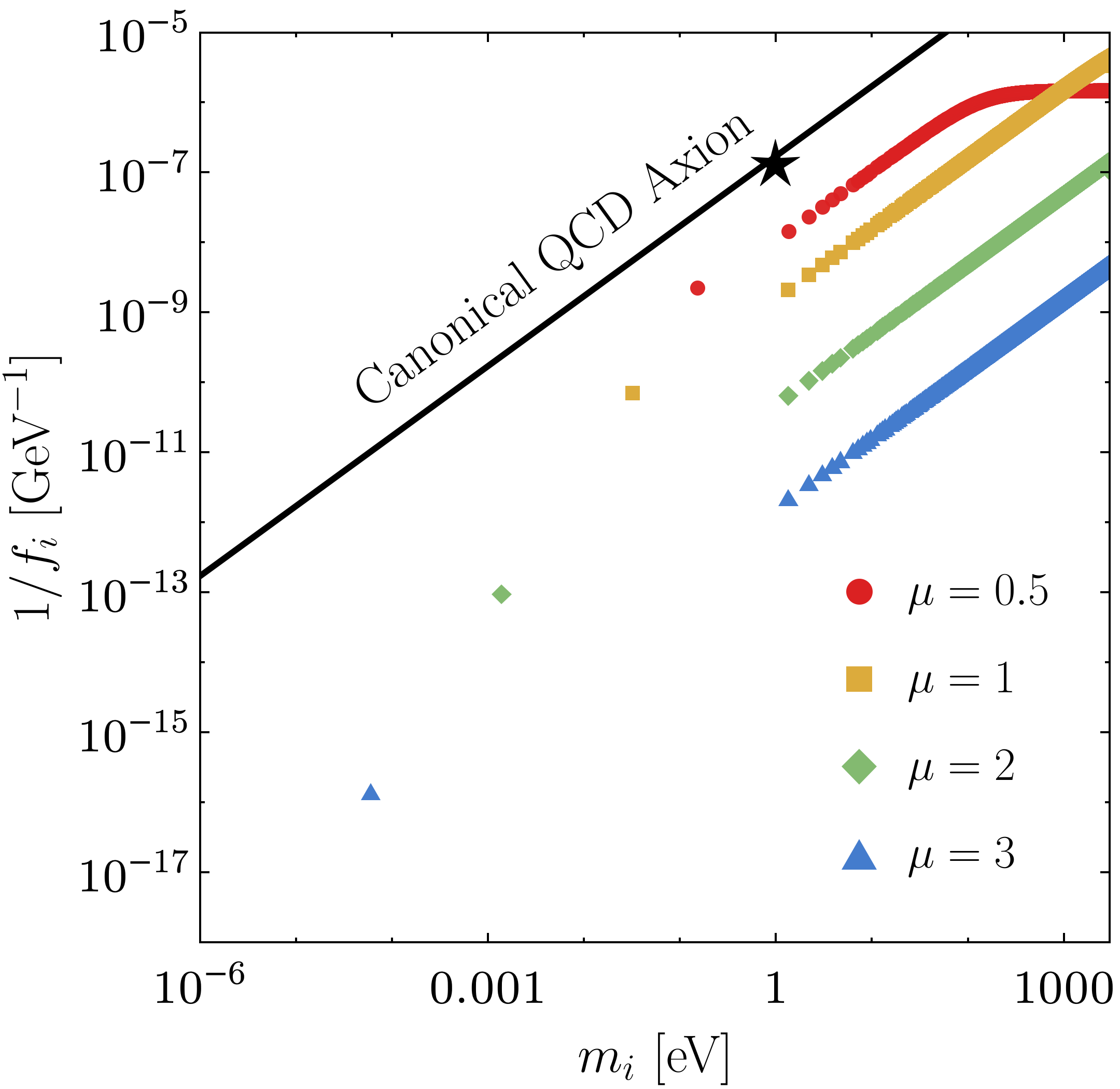}
    \caption{RS ($\mathsf{y}=1$).}
    \label{fig:RS-3}
    \end{subfigure}
    \caption{\it Representative patterns of KK axions in flat (left) and RS (right) models. The solid black line and the black star represent, respectively, the single QCD axion mass-scale relation and the benchmark point of $m_{\rm PQ}=1$\,eV.}
    \label{fig:RS-axions}
\end{figure*}

In the RS model, the computations are more elaborate as $\mu_n \propto \gamma_n$.
The sum in Eq.~\eqref{eq:eigenvalue-eq} can be performed exactly by means of the identity~\cite{Grebenkov_2020}
\begin{equation}
\label{eq:Bessel_1}
    \dfrac{J_{\nu+1}(z)}{J_\nu(z)}=\sum\limits_{k=1}^\infty\dfrac{2z}{\gamma_{\nu,k}^2-z^2}\,,
\end{equation}
where $\gamma_{\nu,n}$ is the $n$-th zero of the Bessel-function $J_\nu(z)$, which leads to the following eigenvalue equation:
\begin{align}
\label{eq:eig-RS}
    &\left(\dfrac{z_1}{2}\right)\dfrac{J_2(z_1)}{J_1(z_1)}=\dfrac{\psi_0^2-\lambda^2}{\psi_1^2}\,,&&z_1\equiv\gamma_1\lambda/\mathsf y\,.
\end{align}
The computation of the $g$-factors can also be performed starting from Eq.~\eqref{eq:Bessel_1}, as
\begin{align}
        \sum\limits_{k=1}^\infty\left(\dfrac{1}{\gamma_{k}^2-z^2}\right)^2 &=\dfrac{1}{2z}\partial_z\left(\dfrac{1}{2z}\dfrac{J_{2}(z)}{J_1(z)}\right) \\
        &=\dfrac{1}{4z^4}\left[z^2-4+\left(z\dfrac{J_0(z)}{J_1(z)}\right)^2\right]\nonumber\,.
\end{align}
By employing the recurrence relation for Bessel functions $J_0(z)/J_1(z) = 2/z - J_2(z)/J_1(z)$ jointly with Eq.~\eqref{eq:eig-RS}, this expression can be further simplified, leading to the RS $g$-factors:
     \begin{align}
              \label{eq:g-RS} 
              g^\text{RS}_\lambda & =\dfrac{\gamma_1^2}{4\mathsf{y}^2}\psi_1^2+\dfrac{2\lambda^2-\psi_0^2}{\lambda^2}+\dfrac{(\lambda^2-\psi_0^2)^2}{\lambda^2}\dfrac{1}{\psi_1^2}\,.
     \end{align}

Let us now study the limiting cases of these formulas, to characterize the axion patterns that emerge in this setup. The lightest eigenvalue can be estimated by Taylor expanding Eq.~\eqref{eq:eig-RS} around $\lambda\sim0$, yielding
\begin{align}
&\label{eq:lightest-RS}
    \lambda_0\approx \dfrac{\sqrt{8}}{\gamma_1}\mathsf y\left(\dfrac{\psi_0}{\psi_1}\right)\,,\\
&\label{eq:G0RS}   g_0\approx\dfrac{\gamma_1^2}{8 \mathsf y^2}\,\psi_1^2\,,\\
    &f_{0} \approx f_4\times\dfrac{\gamma_1^2}{8 \mathsf y^2}\,\left(\dfrac{\psi_1^2}{\psi_0} \right)\,.
\end{align}
In comparison to the flat result, the lightest mode becomes exponentially suppressed by the warp factor, while its $g$-factor and interaction scale are enhanced by $\psi_1^2$.

The mass and couplings of the next $n>0$ modes can be similarly derived; namely in the $\psi_1\gg 1$ limit, one must have $z_1\approx \gamma_{2,n}$ in order to satisfy Eq.~\eqref{eq:eig-RS}. More precisely, we obtain:
\begin{align}
    &\label{eq:mnRS}\lambda_{n>0}\approx\dfrac{\gamma_{2,n}}{\gamma_1} \mathsf y \,,\\
    &g_{n>0}\approx \dfrac{\gamma_1^2}{4 \mathsf y^2}\psi_1^2\,,\\
    &    f_{n>0} \approx 
    f_4\times\dfrac{\gamma_1^2}{2\gamma_{2,n}\mathsf{y}^2}\,\psi_1\,.
\end{align}
We therefore find that several of the modes have the same $g$-factor, mostly identical to that of the zero mode. This confirms the existence of QCD maxion solutions also for this model. Nevertheless, the couplings of these modes are exponentially enhanced relative to that of the zero mode, in deep contrast with what was found in the flat model. Some exact solutions to the equations above are represented in Fig.~\ref{fig:RS-3}, where it is apparent that the distance of the maxions to the QCD axion canonical line is exponentially enhanced by the curvature of the extra dimension.

The coupling of the plateau modes is also exponentially enhanced with respect to the flat model. To see this, we consider $\lambda\sim \gamma_n \mathsf y$ as the QCD contribution is negligible for these modes. The mass of the first mode in the plateau can then be derived by finding the 
$n_\star$ at which the decoupling limit is enforced, or equivalently for which the condition~\eqref{eq:LED-regime} is no longer satisfied.
A more precise value can be obtained by finding the $n_\star$ at which the terms enhanced by the mass in Eq.~\eqref{eq:g-RS} become comparable to the maxions $g$-factor:
\begin{equation}
\label{eq:nstar-RS-estimation}
    \gamma_{n_\star}\approx \pi n_\star \sim  \dfrac{\gamma_1^2}{2\mathsf{y}^2}\,\psi_1^2\,.
\end{equation}
For $n\gtrsim n_\star$, we therefore find:
\begin{align}
&\label{eq:lambda-plateau}
    \lambda_{n,\text{plateau}}\approx \dfrac{\gamma_{n}}{\gamma_1}\mathsf{y}\,,\\
&\label{eq:eq:GRS-plateau} g_\text{plateau}\approx \left(\mathsf{y}\dfrac{\gamma_n }{\gamma_1}\right)^2 \times\dfrac{1}{\psi_1^2}\,,\\
&\label{eq:fplatRS}f_{\text{plateau}} \approx  \dfrac{f_4}{\psi_1}\,.
\end{align}
These results, together with Eq.~\eqref{eq:nstar-RS-estimation}, allow us to estimate the localization of the lightest eigenvalue, as well as of the plateau, for the several scenarios represented in Fig.~\ref{fig:RS-3}. 
We note that a partial study of the mass spectrum of this model was previously presented in the literature~\cite{Flacke:2006re}, while expressions for the masses and some mixing factors were derived in Ref.~\cite{Buyukdag:2019lhh}.

\section{Theoretical and experimental constraints}
\label{sec:constraints}

To infer whether an experiment could detect the KK maxion patterns, it is essential to identify the cutoff of the KK 
EFT in Eq.~\eqref{eq:KKL} below which we can rely on a perturbative treatment of the tower~\footnote{The perturbative unitarity constraints based on the scattering of KK gravitons have been already obtained for the RS model with one extra-dimension, by studying the process $h_i h_j\to h_k h_l$. The corresponding unitarization scale is given by~\cite{deGiorgi:2021xvm}
\begin{equation}
    \label{eq:gravity-cut}\Lambda^\text{gravity}\approx 1.9 \dfrac{M_5}{\psi_1}\approx1.9\left(\dfrac{\mu_1 m_P^2}{\psi_1^2}\right)^{1/3}\,.
\end{equation}
}.
With this aim, let us consider a process with gluons scattering off the gluonic axion combination, at a fixed center of mass energy $\sqrt s \gg m_{n}$. Such a combination interacts with overall strength
\begin{equation}
\frac{1}{F} = \sqrt{\sum_{n=0}^N \left(\frac{\alpha_s\psi_n}{8\pi f_4}\right)^2} \approx \sqrt{N} \,\frac{\alpha_s\psi_1}{8\pi f_4}\,.
\end{equation}
Taking $N\approx \sqrt{s}/\mu_1$, we then expect that the scattering amplitude scales as~\footnote{Note that the sum over propagators can be divided into three regimes, depending on whether $s \gg m_n^2$,  $s \sim m_n^2$ or $s \ll m_n^2$. The resonant regime is expected to involve only a small number of modes while in the third regime, the propagators are dumped by the heavy masses. We, therefore, expect these contributions to be subleading with respect to those in the first regime, where 
the propagator scales as $1/s$ for $N \sim \sqrt{s}/\mu_1$ modes; this justifies the form of the scattering amplitude in Eq.~\eqref{eq:naive-unitarity}.
}
\begin{align}
\label{eq:naive-unitarity}
        \mathcal M & \propto \frac{s}{F^2} \sim  \left(\frac{\alpha_s\psi_1}{8\pi f_4}\right)^2 \frac{s^{3/2}}{\mu_1 }\lesssim 1\,,
\end{align}
from which follows a naive perturbative bound on the energy by requiring that the expansion parameter is less than unity. The result agrees, up to $O(1)$ factors, with the more rigorous calculation presented in App.~\ref{app:perturbative-unitarity} that leads to the following constraint: 
\begin{equation}
 \Lambda\approx (36\pi) \times\left(\dfrac{\mu_1f_4^2}{\psi_1^2}\right)^{1/3}\,.
 \label{eq:Lcutoff}
\end{equation}

By rewriting the equation above in terms of the parameters in the fundamental theory, we find that it is in fact the 5D axion scale that sets the strength of the axion couplings when the~\textit{full} tower is taken into account. This is the expected result of the KK resummation; see Eq.~\eqref{eq:gravity-cut}.

\begin{figure*}[t]
     \centering
    \includegraphics[width=0.8\textwidth]{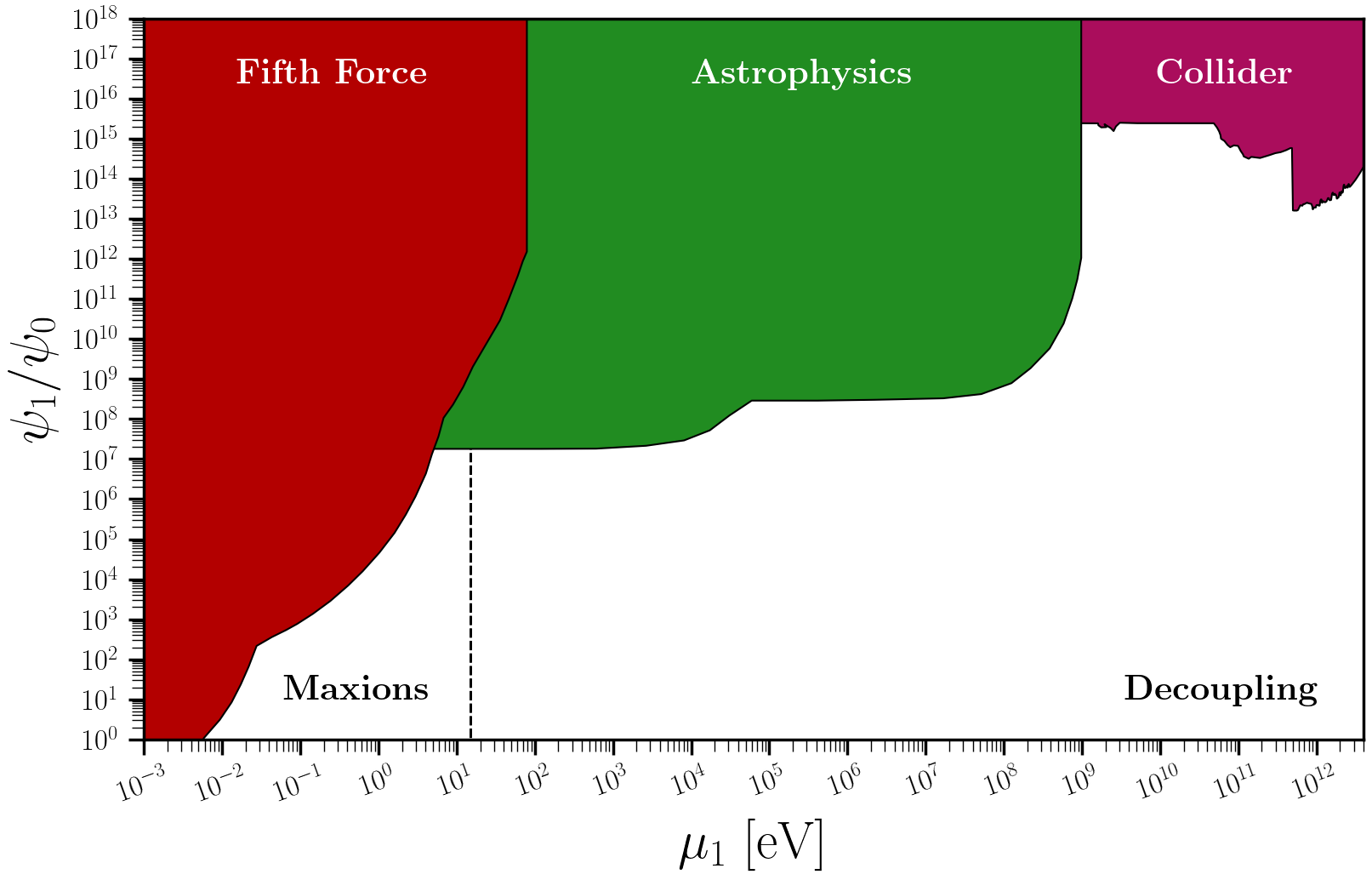}
    \caption{\em Collection of bounds on massive gravitons as a function of the lightest KK graviton mass,~$\mu_1$. 
    The red, green and purple regions are excluded by fifth force experiments~\cite{Hoskins:1985tn,Bordag:2001qi,Mostepanenko:2001fx,Chiaverini:2002cb,Long:2003dx,Chen:2014oda,Tan:2016vwu,Lee:2020zjt}, astrophysics~\cite{Hannestad:2003yd, Cembranos:2017vgi} and collider searches~\cite{deGiorgi:2021xvm,Jodlowski:2023yne,dEnterria:2023npy}, respectively. 
   The black dashed line corresponds to the threshold mass value below which KK maxions can be generated; see Eq.~\eqref{eq:mu-criteria}. This assumes that the KK tower remains perturbative up to TeV energies.
    }
    \label{fig:mix}
\end{figure*}

The final expressions for the $g$-factors will depend on this cutoff scale, as well as on the mass of the lightest graviton. It is therefore necessary to introduce gravity constraints in our setup. A detailed discussion on the impact of the full KK tower to such bounds is beyond the scope of this work and does not change the overall conclusions. We therefore consider the bounds on massive gravity induced only by the presence of the lightest graviton. For very small masses, the latter induces long-distance forces that modify Newton's potential and enter the ballpark of fifth-force searches. Experiments look for deviations
in the form
\begin{equation}
    V(r)=-G_N\dfrac{m_1m_2}{r}\left[1+\alpha e^{-r/\lambda}\right]\,,
\end{equation}
where $G_N$ is the Newton constant.
Taking into account the 4D Planck mass definition in Eq.~\eqref{eq:planck-mass-def} and upon factorizing $G_N$, the contribution from the first massive KK-graviton fixes ${\alpha \approx (4/3) (\psi_1/\psi_0)^2}$ and $\lambda\approx 1/\mu_1$~\cite{Callin:2004py}, so that current limits~\cite{Hoskins:1985tn,Bordag:2001qi,Mostepanenko:2001fx,Chiaverini:2002cb,Long:2003dx,Chen:2014oda,Tan:2016vwu,Lee:2020zjt} can be re-interpreted in terms of our model parameters.
For larger graviton masses, astrophysics bounds from stellar cooling~\cite{Hannestad:2003yd, Cembranos:2017vgi} become dominant, while for masses larger than $\mu_1 \gtrsim 100$~MeV relevant constraints come from beam-dump and collider searches~\cite{deGiorgi:2021xvm,Jodlowski:2023yne,dEnterria:2023npy}. The resulting two-parameter space bounds are presented in Fig.~\ref{fig:mix}. In the flat scenario, where $\psi_1/\psi_0$ is fixed, the bound reads $R^{-1}\gtrsim6\times 10^{-3}$~eV.

To combine these constraints, motivated by the form of the eigenvalue equation~\eqref{eq:eigenvalue-eq}, we consider a simplified approach to extra-dimensional maxions, that relies on using the $2\times 2$ block mass matrix
\begin{equation}
    \begin{split}
        \mathbf{M}^2&=m_\text{PQ}^2\begin{pmatrix}
        (\psi_0)^2 & \psi_0 \psi_1^\pi  \\
        \psi_1^\pi \psi_0  & (\psi_1^\pi)^2+\mathsf{y}^2 \\
    \end{pmatrix}\,.
        \end{split} 
        \label{eq:mass2x2}
\end{equation}
This is equivalent to truncating the sum in Eq.~\eqref{eq:eigenvalue-eq} to the second term.
The lightest eigenvalue can then be predicted up to an accuracy of $\mathcal O(\mu_1/\mu_2)^2$ and, in the limit $\mathsf{y},\psi_0 \ll \psi_1$, reads:
\begin{align}
        \label{eq:lambda022}
    \lambda_{0,2\times 2}^2 &
    \approx \mathsf{y}^2 \left(\dfrac{\psi_0}{\psi_1}\right)^2 \,.\nonumber
\end{align}
In the same limit, we find:
\begin{equation}
    g_0 \approx \dfrac{\psi_1^2}{\mathsf y^2}=\dfrac{\psi_1^2}{f_4^2}\dfrac{\chi_\text{QCD}}{\mu_1^2}\approx (36\pi)^3\dfrac{\chi_\text{QCD}}{\mu_1\Lambda^3}\,, 
    \label{eq:rho}
\end{equation}
which approximates very well the expressions in Eqs.~\eqref{eq:GLED} and~\eqref{eq:G0RS}.
In light of these results, let us argue why this approximation makes sense. The lightest eigenmode obtained in this approach corresponds to the lightest maxion of the full model, while the second one is forced to go close to the canonical line to satisfy the sum rule (for whatever truncation of the mass matrix, the KK interactions preserve always a PQ symmetry at the classical level). By considering more scalar fields i.e. going beyond the $2\times 2$ approximation, the number of maxions increases but the prediction for $g_0$ remains reliable, as the mixing angles between the light fields are all similar.
Nevertheless, the realistic pattern of maxions that reproduce the low-energy consequences of the full model, with no canonical eigenstate in the $\mathsf y\ll 1$ limit, can only be recovered by considering a much larger number of fields $n^\star$, such that $n^\star /g_0 \approx 1$.

In Eq.~\eqref{eq:rho}, we have expressed the $g$-factors in terms of the EFT cutoff identified in Eq.~\eqref{eq:Lcutoff}, up until our perturbative predictions are valid.
After this step, requiring the $g$-factor to be larger than the unity results in the condition
\begin{align}
\label{eq:mu-criteria}
  \mu_1 \lesssim  \left(36 \pi \right)^3 \frac{\chi_{\rm QCD}}{\Lambda^3}  \,,
\end{align}
which is independent of the (exponential) differences between the flat and RS models.
We will consider scenarios with $\Lambda \gtrsim 1$~TeV. Such value ensures that the UV resonances that complete the 5D axion gluonic interaction, such as coloured fermions localized on the IR brane~\cite{Flacke:2006ad}, evade LHC bounds~\footnote{For instance, a KSVZ-like interaction of the form ${S_\Psi \supset \int d^4 x \sqrt{g} \, y_\Psi \overline{\Psi}_L \Phi\Psi_R + \text{h.c.}}$, where $\Phi$ denotes the PQ field, produces heavy fermion masses that get warped down, $m_\Psi^2 \sim y_\Psi^2  A(\pi R)^2 f_5^2$ $\sim y_\Psi^2 \Lambda^3/ (A (\pi R) M_s)$. If $A (\pi R) M_s \sim \Lambda$, the KSVZ fermions are expected at energies of the order of the EFT cutoff scale.}. Even though more exotic constructions with extra branes and throats could weaken these limits, a smaller cutoff scale would induce large corrections to QCD resonances, namely to the mass of the $\eta^\prime$~\footnote{In the limit where $m_{u,d}\to 0$, the correction to the $\eta^\prime$ mass is given by $\Delta (m_\eta^{\prime})^2 \approx N \psi_1^2 \Lambda_{\rm QCD}^4/f_{\rm PQ}^2 \sim \Lambda_{\rm QCD}^4/\Lambda^2$.}. It follows then from the previous expression that non-canonical KK axions could arise for $\mu_1 \ll \mathcal{O}(10)$\,eV. Using the \textit{conservative} 1-graviton bounds in Fig.~\ref{fig:mix}, we find that this value is 3 orders of magnitude above the limit imposed by tests of the fifth force.  

Let us then explore the limiting case of ${\mu_1\sim 10^{-2}}$\,eV which, according to Eq.~\eqref{eq:rho}, would lead to the largest value possible for 
\begin{equation}
\label{eq:g0constraint}
    g_0 \approx 4.5 \times 10^{3} \times \left(\dfrac{10^{-2}\,\text{eV}}{\mu_1}\right)\left(\dfrac{1\,\text{TeV}}{\Lambda}\right)^3\,.
\end{equation}
Such mass is not yet ruled out experimentally in the flat scenario, but for RS setups only larger values for $\mu_1$ are viable and therefore smaller $g$-factors; see Fig.~\ref{fig:mix}. 

The previous expression localizes the maxion modes, but 
those that interact more strongly with the SM are the heavy ones in the plateau, for which
\begin{align}
\label{eq:plateauconstraint}
    \frac{10^ 7\,\text{GeV}}{f_{\text{plateau}}} \approx & \,{1.2} \,{\left(\dfrac{\mu_1}{10^{-2}\,\text{eV}}\right)^{1/2}\left(\dfrac{1\,\text{TeV}}{\Lambda}\right)^{3/2}}\,,
\end{align}
according to Eqs~\eqref{eq:fplateauLED} and \eqref{eq:fplatRS}.  
In terms of the coupling represented in Fig.~\ref{fig:schematic},
\begin{equation}
    g_{a_i\gamma\gamma} \equiv \frac{\alpha}{2 \pi} \frac{|C_{a\gamma}|}{f_i}\,,
\end{equation}
we obtain $g_{a\gamma\gamma\,,\text{plateau}} \approx \,{2.5\times 10^{-10}}\,{\text{GeV}}^{-1}$, after fixing $\mu_1\sim 10^{-2}$\,eV, $\Lambda\sim 1$~TeV and $|C_{a\gamma}| \approx 1.92$ since all axions inherit the same model-independent coupling to photons from the original interaction in Eq.~\eqref{Eq:ActionAxiondDim}. 
Such coupling is on the verge of the constraints imposed by single-axion searches in stars;~see Fig.~\eqref{fig:schematic}.
We note that model-dependent interactions could be introduced in the IR brane to make the KK axions more photophobic, at the cost of some tuning (see e.g.~Ref.~\cite{DiLuzio:2017ogq}). 
Nevertheless, supernova data probes directly the gluonic coupling in Eq.~\eqref{eq:plateauconstraint} for the largest possible displacements identified in Eq.~\eqref{eq:g0constraint}.
Such bound takes into account the production of a single axion in the star, which does not hold in our setup. However, as we show below, the production of additional eigenstates strengthens this limit severely in the parameter space of interest.
\begin{figure*}[t]
     \centering
    \includegraphics[width=0.8\textwidth]{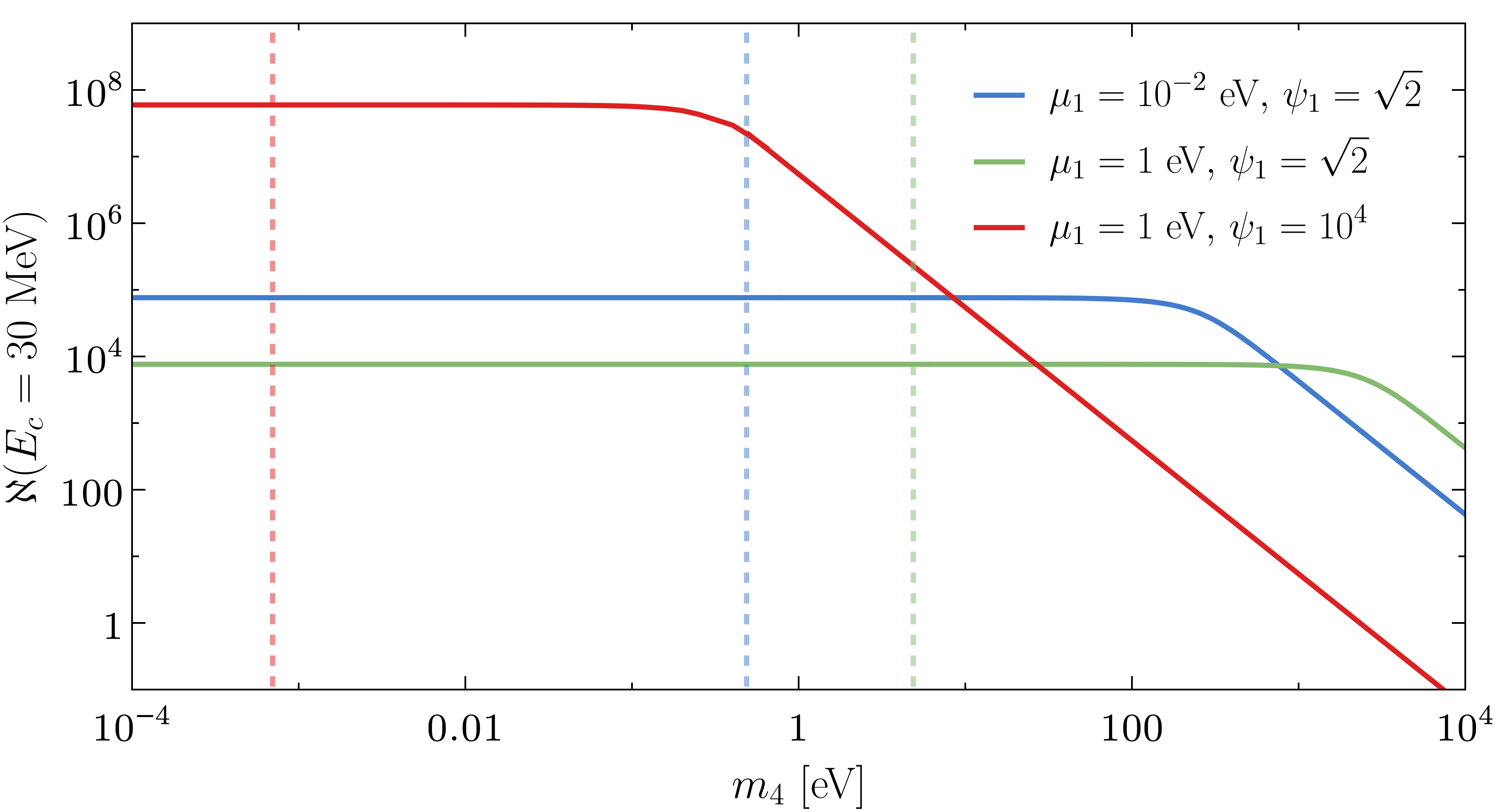}
    \caption{\em The ``rescaling factor" $\aleph(E_c)$, defined in Eq.~\eqref{eq:def-aleph}, as a function of the PQ mass. The different lines correspond to different benchmark values for $\mu_1$ and $\psi_1$ that allow for maxion regimes interpolating the RS and flat scenarios. The characteristic energy was taken to be $E_c=30$\,MeV. The parameter space to the left of the dashed lines, where $\Lambda>1$\,TeV, is consistent with the EFT analysis; see the text for details.
    }
    \label{fig:aleph}
\end{figure*}

By employing the approach of Ref.~\cite{Dienes:2012jb}, one can obtain the effective interaction scale of the KK axions, integrated up to some characteristic energy $E_c$ (around $30$ MeV for typical supernovae core temperatures):
\begin{equation}
    \frac{1}{f_{4,\text{eff.}}}\equiv \frac{\aleph (E_c)}{f_4}\,,
\end{equation}
where
\begin{equation}
\label{eq:def-aleph}
    \aleph^2 (E_c)\equiv f_4^2 \sum\limits_{(\lambda m_4)<E_c} \frac{1}{f_\lambda^2}=\sum\limits_{(\lambda m_4)<E_c} \left(\frac{\lambda^2}{g_\lambda}\right)\,.
\end{equation}
Let us now take some interesting limits of this expression, assuming $E_c\gg \mu_1$. In flat scenarios, we obtain:
\begin{align}
    \aleph_\text{flat}^2 (E_c)\approx 2\times\begin{cases}
     \frac{1}{3\pi^2}\left(\frac{\mu_1E_c^3}{m_4^4}\right) & m_4\gg \mu_1\,,\\
      \left(\frac{E_c}{\mu_1}\right) & m_4\ll \mu_1\,;
    \end{cases}
    \label{eq:Nflat}
\end{align}
while in the warped case,
\begin{align}
    \aleph_\text{RS}^2 (E_c)\approx \begin{cases}
      \frac{4}{3\pi\gamma_1}\left(\frac{\mu_1 E_c^3}{\psi_1^2 m_4^4}\right)  & \psi_1 m_4\gg \mu_1\,,\\
      \frac{\gamma_1}{\pi}\left(\frac{\psi_1^2\,E_c}{\mu_1}\right)  & \psi_1 m_4\ll \mu_1\,.
    \end{cases}
    \label{eq:Ncurved}
\end{align}

The results in Eq.~\eqref{eq:Nflat}, applicable to flat extra-dimensional models, agree with previous findings reported in Ref.~\cite{Dienes:2012jb}. In maxion-dominated regimes, $\aleph_{\rm flat}^2 \sim \mathsf y \lesssim 1$, so that the collective KK bounds can become weaker relative to those obtained for a single axion. In contrast, in the regime dominated by the plateau modes, the relation $1/f_{4,\text{eff.}} \sim \sqrt{N}/ f_4$, where $N\approx E_c/\mu_1$, leads to a substantial strengthening of these constraints. Note that the transition between this mass-dependent and flat regimes occurs approximately at
\begin{equation}
    m_4^\star \approx \frac{\sqrt{\mu_1 E_c}}{2\,\psi_1}\,.
\end{equation}

By extending these results to the warped case, we find that the curvature exponentially suppresses the maxion regimes, while it enhances the plateau contribution. Whichever of these effects dominates, and therefore the fate of the bounds, depends on the parameter space of interest.

Figure~\ref{fig:aleph} shows the values of the ``rescaling factor", $\aleph$, computed numerically for the flat and different RS scenarios. In this figure, we can clearly identify the transition between maxion and plateau dominated regimes, as well as other features discussed previously. 
Choosing $\mu_1 = 10^{-2}$\,eV, which produces the largest $g$-factors allowed by fifth force constraints in the flat scenario (see Eq.~\eqref{eq:g0constraint}), we find that the rescaling factor is dominated by the plateau modes and, therefore, it is extremely large,~$\aleph_{\rm flat} \sim 10^5$.
To compare this with curved models, we have chosen a larger value for $\mu_1 = 1$\,eV, that complies with gravity constraints for $\psi_1$ as large as $10^4$. 
For the same $\Lambda =1$\,TeV (represented by the vertical lines in Fig.~\ref{fig:aleph}), we infer that the warped scenario worsens significantly the bounds, with $\aleph_{\rm RS} \sim \aleph_{\rm flat} \psi_1 \sim 10^{8}$. Such a benchmark corresponds already to a very small displacement of the KK modes from the QCD canonical line, $g_0\sim 10^2$.

We can take the argument further by computing the minimum value of $m_4$, in Eq.~\eqref{eq:Ncurved}, which is compatible with a suppression of the bounds i.e. $\aleph(E_c)<1$. Plugging in this value in the expression for the maxion $g$-factor, we obtain:
\begin{align}
    &m_4^2\gtrsim \left(\frac{\mu_1 E_c^3}{\psi_1^2}\right)^{1/2} &&\implies &&g_0\approx \frac{\psi_1^2}{\mathsf{y}^2}\gtrsim \psi_1 \left(\frac{E_c}{\mu_1}\right)^{3/2}\,.
\end{align}
Written in terms of the cutoff scale, as in Eq.~\eqref{eq:rho}, such condition implies that
\begin{align}
    \mu_1 &\gtrsim \frac{\psi_1^2}{(36\pi)^6}\frac{\Lambda^6 E_c^3}{\chi_\text{QCD}^2}\,\\
    &\nonumber\approx 10^{10}\,\psi_1^2\,\left(\frac{\Lambda}{1~\text{TeV}}\right)^6\left(\frac{E_c}{30~\text{MeV}}\right)^3~\text{GeV}\,,
\end{align}
which is clearly incompatible with the maxion regime. Indeed, the equation above is in strong contradiction with Eq.~\eqref{eq:mu-criteria}. This shows unambiguously that $\aleph(E_c)>1$ even for maxion regimes, unless $E_c$ is comparable to $\mu_1$.

We, therefore, conclude that the maxion patterns from the extra-dimensional models analysed in this work are in severe tension with the joint combination of gravity and astrophysical constraints.

Using this simpler $2\times2$ approximation, we also checked the possibility of generating maxions in the continuum clockwork model~\cite{Giudice:2016yja}. We have found that also for this model the previous conclusions hold: there is no parameter space that complies with gravity bounds where $g_0>1$. This is in part expected as the clockwork mechanism was constructed to suppress mixings between neighbours.

\section{Generalizations}
\label{sec:generalizations}
In the previous discussion, we explored the consequences of only one extra dimension and exploited the common WFs of the KK gravitons and axions. One might then wonder if different conclusions could be drawn in more general scenarios.
Generalizations to the previous assumptions include an extension of the number of extra spacetime dimensions, with the axion propagating only in a subset of those, $\delta^\star<\delta$; and a misalignment between the gravitons and axion WFs due to more complex dynamics of the PQ field in the bulk.

In the following, we briefly discuss such points with some examples.

\subsection{Beyond one extra dimension}
\label{sec:beyond-one}
The consequences of considering a larger number of extra dimensions are not trivial to address and depend on the size of each dimension as well as its compactification. In general, in $\delta$ extra-dimensions, the axion field can be expanded as
\begin{equation}
    a(x,\,\mathbf y)=\dfrac{1}{(2\pi R M_s)^{\delta/2}}\sum\limits_{\vec{n}}a_{\mathbf{n}}(x) \,\psi_{\mathbf n}(\vec{y})\,.
\end{equation}
This results in the generalized KK Lagrangian
\begin{align}
        \mathcal{L}_4&=\sum\limits_\mathbf n\left(\dfrac{1}{2}(\partial_\mu \hat a_\mathbf n)^2- \dfrac{1}{2}\mu_\mathbf n^2 \hat a_\mathbf n^2\right)\nonumber \\
        &+\dfrac{1}{f_4}\dfrac{\alpha_s}{8\pi}\sum\limits_\mathbf n\left(\hat a_\mathbf n \psi_\mathbf n \right)\Tilde{G}_{\mu\nu}G^{\mu\nu}\,,
\end{align}
and in the generalized eigenvalue equation
\begin{equation}
\label{eq:generalised-eigen}
   \,\sum\limits_{|\vec{n}|\leq N}\dfrac{(\psi_{\vec{n}})^2}{\lambda^2-(\mu_{\vec{n}}/\mu_1)^2\mathsf{y}^2} = 1\,;
\end{equation}
see App.~\ref{app:generalization} for details. As before, $\mu_1$ is the lightest mode of the theory, while $N$ denotes the largest KK-number allowed by perturbative unitarity constraints i.e. $N\equiv \Lambda/\mu_1$. 
Note that the above expression is completely general and holds independently of the number of dimensions and type of compactifications. It also shows a new feature compared to the one-dimensional case: the sum diverges logarithmically and exponentially with $N$ for $\delta=2$ and $\delta\geq 3$, respectively. This suggests a stronger dependence of both the eigenvalues and the $g$-factors on the cutoff scale with respect to the one-dimensional case. 

To find explicitly these expressions, let us consider a scenario in which~\textit{each} of the $ \delta$ extra dimensions respects the same orbifold symmetry as considered in the previous sections, so that the WFs can be written as ${\psi_{\mathbf n} \equiv \psi_{n_1}(y_1)\times\dots\times \psi_{n_\delta}(y_\delta)}$.
The smallest eigenvalue in our problem can be then computed
by approximating
\begin{equation}
    \sum\limits_{\vec{n}}f(|\vec{n}|)\approx\int d^\delta n f(|\vec{n}|)=S_{\delta-1}\int\limits_{1}^N dn\, n^{\delta-1} f(n)\,,
\end{equation}
where $S_{\delta-1} \equiv 2\pi^{\delta/2}/\Gamma(\delta/2)$ is the surface area of the $\delta$-ball. Such an approximation is sufficient to capture the relevant behaviour of the expressions with $\delta$ (in particular the leading growth with $N$), but further precision can be achieved by including Euler-Mclaurin correction terms~\cite{polytopes}.
In the limits $\psi_1\gg \psi_0$ and $N\gg 1$, we find~\footnote{The $2^\delta$ stems from summing only over the positive $\vec{n}$.}:
\begin{equation}
    \lambda_0^2 \approx \mathsf{y}^2\left(\dfrac{\psi_0}{\psi_1}\right)^{2\delta}\left(\frac{2^\delta }{S_{\delta-1}}\right)\begin{cases}
       1 & \delta=1\,,\\
       1/\log N & \delta=2\,,\\
     (\delta-2)/N^{\delta-2}   & \delta\geq 3\,.
    \end{cases}
\end{equation}
Notice that in the case $\delta=1$ we recover our previous result, which is independent of the large $N$ behaviour.

The corresponding $g$-factor reads in turn
\begin{equation}
    g_0 \approx \frac{\psi_1^{2\delta}}{\mathsf{y}^2}\left(\frac{S_{\delta-1}}{2^\delta }\right)\times \begin{cases}
       1 & \delta=1\,,\\
       \log N & \delta=2\,,\\
     N^{\delta-2}/(\delta-2)   & \delta\geq 3\,;
    \end{cases}
\end{equation}
and consequently:
\begin{equation}
    f_0 \approx f_4\times \frac{\psi_1^{2\delta}}{\psi_0^\delta}\left(\frac{S_{\delta-1}}{2^\delta \mathsf{y}^2}\right)\times \begin{cases}
       1 & \delta=1\,,\\
       \log N & \delta=2\,,\\
     N^{\delta-2}/(\delta-2)   & \delta\geq 3\,.
    \end{cases}
\end{equation}

The previous expressions show that the potential maxion displacements grow exponentially with the number of dimensions. 
Nevertheless, the unitarity and gravity constraints also become stronger. Indeed, the cutoff scale must be modified by the enhanced coupling and number of modes contributing to the amplitude, as discussed in App.~\ref{sec:PUC}. This results in:
\begin{equation}
   \Lambda^{\delta+2} \approx (36\pi)^3\left(\frac{ 2^\delta\, \delta}{S_{\delta-1}}\right)\left(\frac{\mu_1^\delta f_4^2}{\psi_1^{2\delta}}\right)\,;
\end{equation}
By re-expressing $g_0$ in terms of this scale, we find:
\begin{equation}
    g_0\approx (36\pi)^3 \frac{\chi_\text{QCD}}{\mu_1\Lambda^3} \begin{cases}
      1\,, & \delta=1\,,\\[1ex]
       \left(\dfrac{\mu_1}{\Lambda}\right)\,\log\left(\dfrac{\Lambda^2}{\mu_1^2}\right)\, & \delta=2\,,\\[2ex]
     \left(\dfrac{\delta}{\delta-2}\right)\  \left(\dfrac{\mu_1}{\Lambda}\right)\,   & \delta\geq 3\,.
    \end{cases}
    \label{eq:gdelta}
\end{equation}
The result is quite remarkable. On the one hand, it does not depend on $\delta$ for a large number of extra dimensions. On the other hand, all results beyond $\delta=1$ are suppressed by $\mu_1/\Lambda$. In fact, for $\delta\geq 3$ the dependence on $\mu_1$ completely drops out. Therefore, we conclude that the inclusion of more extra dimensions further precludes the possibility of maxion solutions.

On the experimental side, the contribution of the lightest gravitons with mass $\mu_1$ to fifth force experiments gets rescaled by the number of dimensions, i.e. $\alpha \propto \delta (\psi_1/\psi_0)^{2}$. The lowest values of $\mu_1$ which were allowed in the $\delta=1$ case are consequently excluded in scenarios with more spacetime dimensions. Therefore, if the axion propagates in only a subset $\delta^\star \leq 2$ of the total number of spacetime dimensions, the bounds on $g_0$ will become stronger.

One might still wonder about the consequences of introducing non-universal features among the extra dimensions.
 Such a case is more involved and does not allow a general answer, as fifth-force searches do not constrain the same combination of WFs and masses that enter the eigenvalue equation. Nevertheless, let us imagine a scenario where a series of massive gravitons (and therefore axions) develops much larger WFs than the rest, potentially giving an important contribution to the maxions $g$-factor. While these would contribute to the eigenvalue equation with a factor $\sim (\psi_{\vec{n}^\star}/\mu_{\vec{n}^\star})^2$, this is not the combination bounded by gravity tests, $\sim (\psi_{\vec{n}^\star})^2e^{-\mu_{\vec{n}^\star}r}$.
 In particular, the exponential suppression could effectively screen the contribution of these modes to the Newton potential, such that the corresponding WFs would remain essentially unconstrained by tests of the fifth force. This scenario would then have the potential to change our conclusions concerning the largest value allowed for $g_0$. However, the contribution of such modes to astrophysical probes as well as to the scattering amplitude that determines the unitarization scale is expected to scale in the same way as the contribution to the $g-$factor. Therefore, these two constraints could close the parameter space for maxions even in this case.

Overall, while the above arguments are not sufficient to completely rule out the KK maxion scenarios, they suggest that it is unlikely to find such solutions in a generic extra-dimensional model. 

\subsection{A different VEV profile in the bulk}
The EFT in Eq.~\eqref{Eq:ActionAxiondDim}, studied throughout this work, could be generated by a constant VEV profile of the PQ field on the bulk.
A possible straightforward extension of our setup is then to consider a less trivial, namely $y$-dependent, VEV profile for this field. 
To discuss this case, we write, in a flat spacetime background, a model for a complex PQ field, $\Phi=\rho e^{ia/f_5}/\sqrt{2}$, freely propagating in the bulk:
\begin{equation}
\Scale[0.97]{
    S_\Phi = M_s \int\diff^4{x}\diff{y}\sqrt{g}\,\bigg[\partial_A \Phi^\star \partial^A\Phi 
    -m^2|\Phi|^2\bigg]\,.
    }
\end{equation}
The case of a warped scenario is discussed in Ref.~\cite{Goldberger:1999wh}.
We additionally include a localized potential to induce 
a VEV $\left<\rho(y)\right>_{y=\pi R}=f_5$ on the IR brane, i.e. 
\begin{align}
    S_\text{IR} &\subset  \int\diff^4{x}\diff{y}\sqrt{g}\,\delta(y-\pi R)\bigg[g^{AB}\partial_A \Phi^\star \partial_B\Phi \nonumber\\
    &-\dfrac{\lambda_5}{2}|\Phi|^2(|\Phi|^2-f_5^2)\bigg]\,.
\end{align}

Let us now discuss the $y$-profile of $\rho(x,y)=\left<\rho(y)\right>+\tilde{\rho}(x,y)$. The free EOM for the $\left<\rho(y)\right>$ reads
\begin{align}
    &(\partial_5^2-m^2)\left<\rho(y)\right>=0\,,
\end{align}
where $m^2=\lambda_5f_5^2$ is the mass of the radial mode after spontaneous symmetry breaking.
By imposing the boundary conditions
\begin{align}
    &\partial_5\left<\rho(y)\right>_{y=0}=0\,, & \left<\rho(\pi R)\right>=f_5\,,
\end{align}
we obtain:
\begin{equation}
  \left<\rho(y)\right>=  f_5\times\dfrac{\cosh(my)}{\cosh(m\pi R)}\,.
\end{equation}

In turn, the axion EOM reads
\begin{equation}
   \partial_A\left[\partial^A a\left(\dfrac{\left<\rho(y)\right>}{f_5}\right)^2\right]=0\,.
\end{equation}
By employing the KK decomposition
\begin{equation}
    a(x,y)=\dfrac{1}{\sqrt{2\pi R M_s}}\sum\limits_{n=0}^\infty a_n(x) \phi_n(y)\,,
\end{equation}
we obtain a {modified} Sturm-Liouville equation:
\begin{align}
\label{eq:SL-eq}
     -\partial_5\left[p(y)\partial_5\phi_n(y)\right]=\mu_n^2 r(y)\phi_n\,,
\end{align}
where
\begin{equation}
     r(y)=p(y) \equiv \left(\dfrac{\left<\rho(y)\right>}{f_5}\right)^2\,.
\end{equation}
To find the solutions for $\phi_n$, we require that the derivatives vanish at the boundaries, 
\begin{align}
    &\left(\partial_5 \phi_n(y)\right)_{y=0,\pi R}=0\,.
\end{align}

A constant function is a solution of Eq.~\eqref{eq:SL-eq} for $\mu_0=0$, corresponding to the lightest mode. After normalization~\footnote{As in Eq.~\eqref{eq:norm-conds}, we require:
\begin{equation}
    \frac{1}{2\pi R}\int\limits_{-\pi R}^{\pi R}\text{d}y\,r(y)\,\phi_{i}(y)\phi_j(y)=\delta_{ij}\,.
\end{equation}}, and defining $\nu\equiv mR$, we obtain:
\begin{align}
        \phi_0 & =\sqrt{\dfrac{4\pi\nu \cosh^2(\pi\nu)}{2\pi\nu+\sinh(2\pi\nu)}}  \\
        & =\begin{cases}
            1\,,&m\to0\,;\\
            \sqrt{2\pi m R}\,,&m\to\infty\,.
        \end{cases} \nonumber
\end{align}
The flat model is recovered by taking $m\to 0$.

We turn now into the massive modes, for which we obtain~\footnote{Again, we choose the signs that are convenient in later steps, but with no impact on the results.}:
\begin{equation}
\label{eq:WF-bulk}
    \phi_n(y)=
    (-1)^n  \dfrac{\cos (y  \tilde\mu_n)}{\cosh(m y)} f(\nu)\,,
\end{equation}
with
\begin{equation}
    f(\nu) \equiv \sqrt{\frac{4\pi R \tilde\mu_n\cosh^2(\pi\nu)}{2 \pi  R  \tilde\mu_n+\sin (2 \pi  R  \tilde\mu_n)}}\,.
\end{equation}
The corresponding masses are given by
\begin{equation}
    \mu_n^2=m^2+\tilde\mu_n^2\,,
\end{equation}
where $\tilde\mu_n$ are the solutions to the following equation:
\begin{equation}
\label{eq:mass-condition-bulk}
\Scale[0.92]{    m \tanh (\pi  \nu) \cos (\pi  R \tilde\mu_n)+\tilde\mu_n \sin (\pi  R \tilde\mu_n)=0\,.}
\end{equation}
Such an equation has no closed solution, but in the limit of small and large $m$, one finds: 
\begin{equation}
    \tilde\mu_n=\dfrac{1}{R}\,\begin{cases}
        n\,, & m\to 0\,;\\
        \left(n+1/2\right)\,,&m\to\infty\,,
    \end{cases}
\end{equation}
with $n\in\mathbb{N}$.
In both cases, the masses are bounded from below, $\mu_n\geq m$. 

By making use of Eq.~\eqref{eq:mass-condition-bulk}, we can write the WFs at $y=\pi R$ as
\begin{equation}
    \phi_n(\pi R)=
    -(-1)^n \dfrac{\tilde\mu_n}{m}\dfrac{\sin ( \pi R\tilde\mu_n)}{\sinh(\pi\nu)} f(\nu) \,,   
\end{equation}
which in the previously considered limits read:
\begin{equation}
    \phi_n(\pi R)=\sqrt{2}\,\begin{cases}
        1\,, & m\to 0\,,\\
        -\left(\dfrac{\tilde\mu_n}{m}\right)\,, & m\to\infty\,.
    \end{cases}
\end{equation}

Using these results, it becomes clear that in the presence of a massive bulk PQ field, the KK axions become heavier and the WFs more suppressed, for $n>0$. The corresponding mass matrix is therefore expected to become more diagonal, and the $a_{n>0}$ eigenstates more decoupled from the zero mode. In this way, in the limit where the predictions are distinct from the ones analysed in the previous sections, this scenario is expected to deliver a canonical QCD axion.

\section{The only possible pattern}

We have concluded that non-canonical QCD axion patterns induced by a PQ bulk field are in significant tension with fifth force and astrophysical bounds. This claim was verified in both flat and RS scenarios, for different VEV profiles of the bulk field, and independently of the number of extra orbifolded spacetime dimensions. 
In light of these results, we therefore expect that only canonical KK patterns emerge in these generic scenarios.

The canonical pattern consists on having one axion in the canonical QCD band plus a plateau of heavy modes, located a distance ${\Delta m = \mu_1}$ away from the former. Even though the plateau modes are expected to be as weakly interacting as the QCD axion, their resummation could make the pattern more visible to experiment and eventually lead to the identification of the QCD axion interaction scale.
For instance, in a broadband experiment like CAST, the overall effect of the KK exchange induces a stronger coupling to photons given by
    \begin{align}
        g_{a\gamma\gamma,\,\text{eff}}^2 & \sim N g_{a\gamma\gamma}^2 \\
        & \approx N \left(\frac{\alpha}{2 \pi}\right)^2 \left(\frac{\psi_1}{f_4}\right)^2 \nonumber\\
        & \approx \left(\frac{\text{keV}}{\Delta m}\right) \left(\frac{\alpha}{2 \pi}\right)^2 \left(\frac{1}{f_{\rm{plateau}}}\right)^2 \nonumber\,,
    \end{align}
which illustrates how a discovery of a multiple KK signal could translate into an identification of the zero-mode in the canonical QCD axion band.
Additional implications of the KK photon coupling have been discussed, for instance, in Refs.~\cite{Dienes:1999gw, Horvat:2003xd, Bastero-Gil:2021oky, NEWS-G:2021vfh}. It is also worth pointing out that, while in these regimes the massive KK modes would be decoupled from the solution to the strong CP problem, they could contribute sizably to the dark matter abundance of the Universe, affecting significantly the predictions with respect to the single axion case~\cite{Dienes:2011ja, Dienes:2011sa, Dienes:2012jb, Buyukdag:2019lhh}. 

For large values of $\mu_1\sim\mathcal{O}(\text{MeV})$, such that the second mode would appear at LHC scales, the resummation effects are expected to be substantially weaker, and instead only a small number of modes could be tested resonantly, with no significant implications to the QCD axion. (This conclusion could be substantially different, however, in models capable of generating large displacements of KK maxions.)

Finally, let us also comment on which UV parameters could generate the patterns discussed in this work.
As we showed in previous sections, the mixing among KK modes is fully predicted by the compactification of the higher-dimensional theory, and it is controlled by the mixing parameter
\begin{equation}
    \frac{\psi_1}{\mathsf{y}} \sim \psi_1^2 \frac{\Lambda_{\rm QCD}^2 m_P}{(M_5 f_5)^{3/2}} \,,
    \label{eq:UV1}
\end{equation}
where, in the second step, we assumed that ${M_s \sim f_5 \leq M_5}$. In particular, if we further require that all scales are equal in the UV model, it follows that 
\begin{equation}
    \frac{\psi_1}{\mathsf{y}} \sim \left( \frac{\Lambda_{\rm QCD}}{A(\pi R) M_5} \right)^2 \ll 1\,,
\end{equation}
due to the requirement of perturbative unitarity in the gravity sector up to the TeV scale (see Eq.~\eqref{eq:gravity-cut}). The most natural extra-dimensional setups hence lead to the decoupling of the massive modes, and consequently to the presence of a single axion in the canonical QCD line. More hierarchical scenarios, with $M_5 \gg f_5$, would then be required to produce large mixings among the KK modes.

Such hierarchical scenarios could nevertheless be realized in Nature, at the cost of raising new questions e.g. related to the mechanism to stabilize the different UV scales. The interest of our results is that we proved, independently of relations among the fundamental parameters, that all the exotic scenarios which could arise from the UV theory are phenomenologically constrained.

Although this holds in generic models, established upon standard KK WFs and compactifications, engaging in more elaborate constructions to allow sizable KK mixing could have profound consequences on the axion phenomenology. It remains to be tested, for instance, the possibility to generate the extra-dimensional maxion patterns in:
\begin{enumerate}
    \item Scenarios where QCD also propagates in the bulk of the extra dimensions $\delta$~\cite{Gherghetta:2020keg,Bedi:2024wqg}, enhancing a $\delta^\star$-axion mass and therefore that of the 4D KK modes\,;
    \item 
    More exotic constructions where the WFs of the axion are disentangled from gravity bounds, e.g. due to the combination of additional bulk fields and more involved compactifications;
    \item The string axiverse, which provides extra mass sources for the KK modes of higher dimensional fields, in setups where the PQ symmetry remains essentially unbroken at low energies. Under the assumption that the different instanton scales are highly hierarchical, it has been found that the mixing of light ALPs with the axion gluonic combination is very weak~\cite{Gendler:2023kjt}. 
    However, a study of more aligned regimes, that could lead to the exotic phenomenology discussed in this work, is still lacking.
\end{enumerate}
We plan to explore some of these directions in the near future.

\begin{acknowledgments}
The authors would like to thank Belen Gavela for the insightful discussions that helped shape the direction of this work. We are also grateful to Javi Lizana and Pablo Quilez for the enlightening discussions, and to Enrique Fernandez, Pierluca Carenza, Jakob Moritz, and Javi Serra for their valuable input on specific aspects of this study. We also thank Peter Cox for the interesting comments on the manuscript and our referee, for pointing out relevant literature that was overlooked in the first version of this work.
The authors would also like to thank Joerg Jaeckel and the Institute for Theoretical Physics of the University Heidelberg for the warm hospitality during part of the realization of this work. The work of AdG was supported by the European Union’s Horizon 2020 Marie Sk\l odowska-Curie grant agreement No 860881-HIDDeN and the STFC under Grant No. ST/X003167/1. This article/publication is based upon work from COST Action COSMIC WISPers CA21106, supported by COST (European Cooperation in Science and Technology).
\end{acknowledgments}

\bigskip
\appendix

\section{Eigensystem of the mass matrix}
\subsection*{Eigenvalues in one extra-dimension}
\label{app:eigenvalues}
To obtain the eigenvalues in our problem, we must compute the determinant of the following $(N+1)^2$ matrix, that accounts for the mixing of the zero mode with $N$ massive modes:
\begin{align}
\label{eq:form-onedim}
    \mathbf{A} & \equiv \frac{\mathbf{M}^2}{m_{\rm PQ}^2} - \lambda^2 \mathbf{1}  \\
    & =\begin{pmatrix}
        c_0 & \zeta & \zeta &\dots&\zeta\\
        \zeta & \sigma+c_1 & \sigma & \dots&\sigma\\
        \zeta & \sigma &\sigma+ c_2 & \dots&\sigma\\
        \dots &  \dots &    \dots&\dots&  \sigma\\
        \zeta & \sigma& \sigma &\sigma &\sigma+ c_N
    \end{pmatrix}\,, \nonumber
\end{align}
    where $\zeta \equiv \psi_0 \psi_i$, $\sigma\equiv (\psi_i)^2$, $c_0 \equiv (\psi_0)^2 - \lambda^2$ and $c_i\equiv (\mu_i/\mu_1)^2\mathsf{y}^2-\lambda^2$. %
    By making use of the fact that linear combinations of rows ($\ell$) and columns ($c$) do not change the determinant, we can greatly simplify this matrix. For instance, by making $\ell_{n}\to \ell_{n} - \ell_{n+1} $, starting from the second row i.e. $n=1$, we find: 
\begin{equation}
    \det(\mathbf{A})=\begin{vmatrix}
        c_0 & \zeta & \zeta &\zeta &\dots&\zeta\\
        0 & c_1 & -c_2 & 0 &\dots& 0\\
        0 & 0 & c_2 & -c_3 & \dots&0\\
        0 & 0 &0&  c_3& \dots&0\\
        \dots &  \dots &    \dots&\dots&\dots& -c_N\\
        \zeta & \sigma& \sigma &\sigma&\sigma &\sigma+ c_N
    \end{vmatrix}\,.
\end{equation}
The matrix is now triangular, up to the last line. Assuming $c_i\neq 0$, we will now subtract each of the lines above in order to eliminate consecutively the elements $A_{Nn}$.
For example, after the first iteration $\ell_N\to \ell_N -  (\zeta/c_0)\ell_0$, we get to:
\begin{widetext}
\begin{equation}
    \det(\mathbf{A})=\begin{vmatrix}
        c_0 & \zeta & \zeta &\zeta &\dots&\zeta\\
        0 & c_1 & -c_2 & 0 &\dots& 0\\
        0 & 0 & c_2 & -c_3 & \dots&0\\
        0 & 0 &0&  c_3& \dots&0\\
        \dots &  \dots &    \dots&\dots&\dots& -c_N\\
       0  & -s_1\equiv\sigma-\zeta^2/c_0& -s_1&-s_1&-s_1&c_N-s_1
    \end{vmatrix}\,.
\end{equation}
\end{widetext}
After the second iteration $\ell_N\to \ell_N +  (s_1/c_1)\ell_1$, we can eliminate {$A_{N1}$} so that the next element {$A_{N2}$} becomes $-s_2 \equiv -s_1 - (c_2/c_1) s_1$. Denoting by $-s_n$ the {$A_{Nn}$} term obtained after each iteration, we can work out the recursive relation
\begin{equation}
    s_{n+1}=s_1+\dfrac{c_{n+1}}{c_n}s_n\,,
    \label{eq:s1}
\end{equation}
with the boundary condition
\begin{equation}
    s_N=s_{1}+\dfrac{c_{N}}{c_{N-1}}s_{N-1}-c_N\,.
    \label{eq:sN}
\end{equation}
Using now Eq.~\eqref{eq:s1} to replace the element $s_{N-1}$ in the Eq. above, and doing so consecutively for the elements $s_{N-2}$, $s_{N-3}$, etc., we finally obtain:
\begin{equation}
    -s_N=c_N\left[1-s_1\,\sum\limits_{n=1}^N\dfrac{1}{c_n}\right]\,.
\end{equation}
As the matrix is now upper triangular, the determinant reads
\begin{equation}
    \det(A)=\left(\prod\limits_{n=0}^{N-1}c_n\right) \times (-s_N)\,.
\end{equation}
The eigenvalues are to be found in the zeros of such determinant. By assumption, $c_i\neq 0$, so we must require
\begin{equation}
 0 \overset{!}{=}  s_N\propto1-s_1\,\sum\limits_{n=1}^N\dfrac{1}{c_n} \,, 
 \label{eq:sN0}
\end{equation}
that is, the eigenvalues satisfy the equation
\begin{equation}
    s_1\,\sum\limits_{n=1}^N\dfrac{1}{c_n} = 1\,.
    \label{eq:s1final}
\end{equation}
In the original notation, such a relation reads
\begin{equation}
     \lambda^2\,\sum\limits_{n=1}^N\dfrac{(\psi_n)^2}{(\mu_n/\mu_1)^2\mathsf{y}^2-\lambda^2} = (\psi_0^2-\lambda^2)\,.
\end{equation}
In more compact notation, it follows that
\begin{equation}
\label{eq:U-psi}
   \,\sum\limits_{n=0}^N\dfrac{(\psi_n)^2}{\lambda^2-(\mu_n/\mu_1)^2\mathsf{y}^2} = 1\,,
\end{equation}
where $\mu_0=0$.
\subsection*{Eigenvectors in one extra-dimension}
\label{app:eigenvectors}
We focus now on the eigenvectors {$\vec{u}$} of $\mathbf{M}^2$. As the linear system that we are trying to solve is homogeneous, {$\mathbf{A}\vec{u}=0$}, we can use as a starting point the result found in the previous section, i.e. the matrix
\begin{equation}
\begin{pmatrix}
        c_0 & \zeta & \zeta &\zeta &\dots&\zeta\\
        0 & c_1 & -c_2 & 0 &\dots& 0\\
        0 & 0 & c_2 & -c_3 & \dots&0\\
        0 & 0 &0&  c_3& \dots&0\\
        \dots &  \dots &    \dots&\dots&\dots& -c_N\\
       0  & 0& 0&0&0&-s_N
    \end{pmatrix}\,.
\end{equation}
Let us consider the family of $N+1$ eigenvectors, each labelled by its eigenvalue $\lambda$, with components $u_i$.
All entries in the last row vanish once we employ the condition in Eq.~\eqref{eq:sN0}.
This grants us the freedom to set one component to an arbitrary constant, e.g. $u_N$.
By solving the eigenvector equation, we then find a recursive relation for the eigenvector components
\begin{equation}
    c_i u_{i} =  c_{i+1} u_{i+1}\,,\quad 1\leq i\leq N-1\,,
\end{equation}
that implies:
\begin{equation}
    u_i = \dfrac{c_N}{c_i} u_{N}\,, \qquad \forall i\in[1,N-1]\,.
    \end{equation}
Consequently, the first component can be written as
\begin{widetext}
\begin{equation}
    u_0 =-\dfrac{\zeta}{c_0}\left(\sum\limits_{n=1}^N u_n\right)=-\zeta\dfrac{c_N}{c_0}\left(\sum\limits_{n=1}^N \dfrac{1}{c_n}\right)u_N=-\dfrac{\zeta}{s_1}\dfrac{c_N}{c_0}u_N=\dfrac{\zeta}{\sigma c_0-\zeta^2} c_N u_N\,,
\end{equation}
\end{widetext}
where in the last steps we made use of Eq.~\eqref{eq:s1final}. 

Altogether, the eigenvectors can be expressed as
\begin{equation}
    \Scale[0.85]{\vec{u}=\mathcal{N}_\lambda\left(\dots,\, \dfrac{\psi_1}{(\mu_i/\mu_1)^2\mathsf{y}^2-\lambda^2},\,\dots,\, \dfrac{\psi_1}{(\mu_N/\mu_1)^2\mathsf{y}^2-\lambda^2}\right)^T\,,}
\end{equation}
with the normalization constant $\mathcal{N}_\lambda$ given by
\begin{equation}
    \Scale[1]{\mathcal{N}_\lambda=\left[
    \sum\limits_{n=0}^N\left(\dfrac{\psi_n}{(\mu_n/\mu_1)^2\mathsf{y}^2-\lambda^2}\right)^2\right]^{-1/2}}\,.
\end{equation}
More compactly, we can write
\begin{equation}
\label{eq:final-eigens}
    u_i =\mathcal{N}_\lambda\,\dfrac{\psi_i}{(\mu_i/\mu_1)^2\mathsf{y}^2-\lambda^2}\,.
\end{equation}


\subsection*{Eigenvalues and eigenvectors in arbitrary dimensions}
\label{app:generalization}

The formulas for the eigenvalues and eigenvectors found in the one-dimensional case can be generalized to the case of an arbitrary number of dimensions $\delta$, without the need to make any assumptions on $\psi_\vec{n}$. The key observation is that any matrix with different WF factors can be transformed into a universal one, with non-universal terms only in the diagonal. This allows us to use the previously obtained results.

Let us begin by revisiting the mass matrix structure for $\delta=1$:
\begin{equation}
    \begin{split}
        \left(\mathbf{M}^2\right)_{ij}&=m_\text{PQ}^2\left[\psi_i \psi_j +\mathsf{y}^2\left(\dfrac{\mu_i}{\mu_1}\right)^2\delta_{ij}\right]\,.
        \end{split} 
\end{equation}
The same structure holds for an arbitrary $\delta$ since each vector label $\vec{n}=(n_1,\,n_2,\,\dots,\,n_\delta)$ can be counted using a single $n$ label; e.g. for $\delta=2$:
\begin{align}
    &(0,0)\to n=0\,, && (1,0)\to n=1\,, && (0,1)\to n=2\,,\nonumber\\
    & (2,0)\to n=3\,, && (1,1)\to n=4\,, &&\dots\,
\end{align}
So in this example, the mixing in the entry $\mathbf M^2_{12}$ is given by $\psi_1 \psi_2\equiv \psi_{(1,0)}\psi_{(0,1)} $ in the notation above.
We will identify $\mu_1$ with the mass of the lightest mode in the theory (which now can have several copies).

We now proceed with the computation of the eigenvalues and eigenvectors.
\paragraph{Eigenvalues.}
The eigenvalues are found by requiring that the determinant of the following matrix
\begin{equation}
\label{eq:delta-dim-matrix}
\Scale[0.97]{    
        \mathbf A_{ij}  \equiv m_\text{PQ}^2\left[\psi_i \psi_j +(\mathsf{y}^2(\mu_i/\mu_1)^2-\lambda{^2})\delta_{ij}\right]\,.
    }
\end{equation}
is zero. 
By multiplying a row or a column by a number $\alpha$, the determinant also gets multiplied by $\alpha$. This does not affect the zeros of the characteristic polynomial. We therefore multiply each $i$-row by $\psi_0/\psi_i$ and each $j$-column by $\psi_0/\psi_j$, and proceed to evaluate the equation
\begin{equation}
        \left| \psi_0^2 +\left[(\mathsf{y}^2(\mu_i/\mu_1)^2-\lambda{^2})\frac{\psi_0^2}{\psi_i \psi_j}\right]\delta_{ij}\right| = 0\,.
\end{equation}
This matrix has the same form as the one studied in the one-dimensional case (see Eq.~\eqref{eq:form-onedim}), with constant values in all entries besides the diagonal. Therefore, by applying the same steps as in Sec.~\ref{app:eigenvalues}, we obtain the straightforward generalization of Eq.~\eqref{eq:U-psi}:
\begin{equation}
   \,\sum\limits_{n=0}^N\dfrac{(\psi_n)^2}{\lambda^2-(\mu_n/\mu_1)^2\mathsf{y}^2} = 1\,,
\end{equation}
or equivalently
\begin{equation}
   \,\sum\limits_{|\vec{n}|\leq N}\dfrac{(\psi_{\vec{n}}^\pi)^2}{\lambda^2-(\mu_{\vec{n}}/\mu_1)^2\mathsf{y}^2} = 1\,.
\end{equation}

\paragraph{Eigenvectors.}
The computation of the eigenvectors follows a similar reasoning. We aim to compute the solutions of the homogeneous linear system defined by Eq.~\eqref{eq:delta-dim-matrix}. Being homogeneous, we can multiply each row by a constant without modifying the solutions: we choose $\psi_0/\psi_i$. Regarding the columns, we redefine the $j$th component of the eigenvectors (that is just a variable in our linear system of equations) as
\begin{equation}
    u_j = \frac{\psi_0}{\psi_j}u'_j\,.
\end{equation}
After these replacements, the system is again in the form of the one we analysed in the one-dimensional case. 
This allows us to use the same results derived before, upon replacing, in the end, each $j$th component by $u_j$. By following this procedure, one finds the result formally reads the same and one can replace the sum over the single label $n$ with the sum over all possible vector labels $\vec{n}$.

Finally, this also implies that the expression for the $g$-factors remains formally the same.

\section{The QCD axion sum rule in extra dimensions}
\label{app:differential-g-factors}

The QCD axion sum rule was proved in all generality, for an arbitrary dimensional
mass matrix, in Ref.~\cite{Gavela:2023tzu}. Since the extra-dimensional models considered here are PQ invariant, the sum rule must also apply here. One can check it explicitly using the results derived in the previous section.

With this aim, let us consider the following unit vector:
\begin{equation}
    \Vec{s}=(1,0,0,0,\dots)\,,
\end{equation}
By applying this vector to the squared mass matrix in the original basis, we can extract its first component:
\begin{equation}
   \psi_0^2 =(\mathbf{M}^2)_{00}= \vec{s}^T\mathbf{M}^2\Vec{s}=\vec{v}^T\mathbf{M}^2_\text{phy}\Vec{v}\,.
   \label{eq:twobases}
\end{equation}
This must be equivalent to acting with ${\Vec{v} = U^T \Vec s}$ on $\mathbf{M}^2_\text{phy}$, the physical mass matrix obtained after diagonalization via the rotation matrix ${U=(\mathbf{u}_{\lambda_0},\,\mathbf{u}_{\lambda_1},\,\dots,\,\mathbf{u}_{\lambda_N})}$.
In particular, the components of such a vector are given by:
\begin{equation}
    v_i=\sum\limits_jU_{ji}s_j=U_{0i}\,.
\end{equation}
Using Eq.~\eqref{eq:final-eigens}, it then follows that
\begin{equation}
     \vec{v}^T\mathbf{M}_\text{phy}^2\Vec{v} = \sum\limits_\lambda v_\lambda^2 \lambda^2=\psi_0^2\sum\limits_{\lambda}\dfrac{\mathcal{N}_\lambda^2}{\lambda^2}\,,
\end{equation}
which -- in order to match Eq.~\eqref{eq:twobases} -- requires
\begin{equation}
    \sum\limits_{\lambda}\left(\dfrac{\mathcal{N}_\lambda^2}{\lambda^2}\right) = \sum\limits_{\lambda}\left(\dfrac{1}{g_\lambda}\right)=1\,.
\end{equation}

\section{Perturbative unitarity constraints}\label{sec:PUC}
We will focus separately on the one- and multi-extra-dimensional cases.
\label{app:perturbative-unitarity}
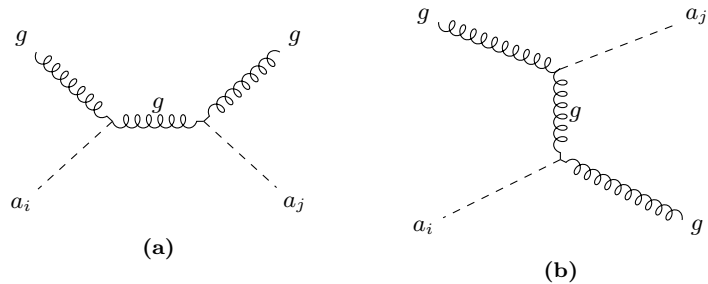
\begin{figure*}[t]
\centering
\begin{subfigure}[c]{0.32\textwidth}
    \centering
\begin{tikzpicture}
  \begin{feynman}
    \vertex (a) {\(g\)};
    \vertex at ($(a) + (1.2cm, -1.1 cm)$) (b);
    \vertex at ($(a) + (0cm, -2.2 cm)$) (c) {\(a_i\)};
    \vertex at ($(b) + (1.2cm, 0cm)$)(d) ;
    \vertex at ($(d) + (1.2 cm, 1.1cm)$) (h1) {\(g\)};
    \vertex at ($(h1) + (0cm, -2.2cm)$) (h2) {\(a_j\)};
    
    \diagram* {
      (a) -- [gluon] (b);
      (b) -- [scalar] (c),
      (b) -- [gluon, edge label=\(g\)] (d);
      (h1) -- [gluon] (d);
      (d) -- [scalar] (h2),
    };
  \end{feynman}
\end{tikzpicture}
\caption{}
\label{fig:FD1}
\end{subfigure}
\begin{subfigure}[c]{0.32\textwidth}
\centering
\begin{tikzpicture}
  \begin{feynman}
    \vertex (a) {\(g\)};
    \vertex at ($(a) + (1.8cm, -0.7 cm)$) (b);
    \vertex at ($(b) + (0cm, -1.2 cm)$) (d);
    \vertex at ($(b) + (1.8cm, 0.7cm)$)(c) {\(a_j\)};
    \vertex at ($(a) + (0 cm, -2.8cm)$) (h1) {\(a_i\)};
    \vertex at ($(c) + (0cm, -2.8cm)$) (h2) {\(g\)};
    
    \diagram* {
      (a) -- [gluon] (b);
      (b) -- [scalar] (c),
      (b) -- [gluon, edge label=\(g\)] (d),
      (d) -- [scalar] (h1);
      (h2) -- [gluon] (d),
    };
    \end{feynman}
\end{tikzpicture}
\caption{}
\label{fig:FD2}
\end{subfigure}
\caption{\em Feynman diagrams contributing to $a_ig\to a_j g$.}
\label{fig:FD-unitarity}
\end{figure*}

Setting $\delta = 1$ at first,
let us consider the elastic scattering $a_ig\to a_ig$ to infer a constraint from perturbative unitarity~\footnote{Note that by considering the $a_ig\to a_ig$ process convoluted with the gluon PDFs, or even other processes involving weaker tree-level couplings of the axions, the unitarity bound would only become weaker.}. The diagrams contributing to the process are depicted in Fig.~\ref{fig:FD-unitarity}. 
We employ partial-wave analysis (see e.g. Ref.~\cite{Chanowitz:1978uj}) and compute the $J=1$ wave in the high energy limit
\begin{equation}
    \mathcal{M}_1\approx\dfrac{1}{32\pi}\int\limits\text{d}(\cos\theta)\,d_{s_1s_2}^{J=1}(\theta)\mathcal{M}_{s_1s_2}(s,\theta)\,,
\end{equation}
where $d_{s_1s_2}^J(\theta)$ is the Wigner d-function, and in this case $\mathcal{M}_{s_1s_2}$ is the amplitude relative to the helicities $s_{1,2}=\pm$ of the initial and final gluons.
The scattering matrix element is non-vanishing for the same initial and final colour state and in the high-energy limits reads
\begin{equation}
    \mathcal{M}_{1,ii}=\begin{pmatrix}
        \mathcal{M}_{1,++}&\mathcal{M}_{1,+-}\\
        \mathcal{M}_{1,-+}&\mathcal{M}_{1,--}
    \end{pmatrix}=\left(\dfrac{\alpha_s \psi_i}{8\pi f_4}\right)^2\dfrac{s}{6\pi}\begin{pmatrix}
        1/4&1\\
        1&1/4
    \end{pmatrix}\,.
\end{equation}
Perturbative unitarity imposes the smallest of the eigenvalues of such matrix to be smaller than $1$, thus implying
\begin{equation}
\label{eq:unitarity-1}
    \left(\dfrac{\alpha_s \psi_i}{8\pi f_4}\right)^2\left(\dfrac{5}{24\pi}\right)s\leq 1\,.
\end{equation}
If we now consider the full set of processes $a_ig\to a_j g$, the number of amplitudes increases. The full scattering amplitude matrix for this set of processes can be block-built starting from $\mathcal{M}_{1,ii}$ such that
\begin{equation}
     \left(\mathcal{M}_{1}\right)_{ij}= \mathcal{M}_{1,ij}\,.
\end{equation}
If we consider the first $N$ axions, then the largest eigenvalue of such matrix is found to grow linearly with $N$ and the condition of Eq.~\eqref{eq:unitarity-1} becomes
\begin{equation}
\label{eq:pert-onedim}
    N\times \left(\dfrac{\alpha_s \psi_i}{8\pi f_4}\right)^2\left(\dfrac{5}{24\pi}\right)s\approx \left(\dfrac{\alpha_s \psi_i}{8\pi f_4}\right)^2\left(\dfrac{5}{24\pi}\right)\dfrac{s^{3/2}}{\mu_1}\leq 1\,,
\end{equation}
where in the last step we approximated again the number of available axions as $N\approx \sqrt{s}/\mu_1$.  The result matches the naive result of Eq~\eqref{eq:naive-unitarity} corrected by a numerical factor of $5/(24\pi)\approx 1/(5\pi)$.

We can therefore identify the cut-off of our EFT to be
\begin{equation}
    \Lambda\approx \left(\dfrac{320\pi^3}{\alpha_s^2}\right)^{1/3}  \left(\dfrac{\mu_1f_4^2}{\psi_i^2}\right)^{1/3}\approx (36\pi)\,\left(\dfrac{\mu_1f_4^2}{\psi_i^2}\right)^{1/3}\,,
    \label{eq:Lambda}
\end{equation}
where we employed $\alpha_s(1\text{TeV})\approx 0.08$~\cite{ATLAS:2023tgo}.

When dealing with a larger number of extra spacetime dimensions, $\delta > 1$, the previous discussion needs to be generalized. We will obtain now the more general expression for the unitarization scale, assuming that all dimensions respect 
the same orbifold symmetry as 
discussed in Sec.~\ref{sec:beyond-one}. 

In such case, Eq.~\eqref{eq:pert-onedim} is modified since the number of modes contributing to the amplitude now grows exponentially.  Denoting by $N$ the number that saturates the condition $\mu_{\vec{n}}\lesssim \sqrt{s}$, the effective number of axions can be estimated as
\begin{equation}
    N_\text{eff}=\sum\limits_{|\vec{n}|\leq N}1\approx \frac{S_{\delta-1}}{2^\delta \delta}N^\delta\,.
\end{equation}
As before, $N\approx \sqrt{s}/\mu_1$ so that Eq.~\eqref{eq:pert-onedim} becomes
\begin{equation}
\label{eq:pert-moredims}
    N_\text{eff}\times \left(\dfrac{\alpha_s \psi_1^\delta}{8\pi f_4}\right)^2\left(\dfrac{5}{24\pi}\right)s\leq 1\,,
\end{equation}
leading to
\begin{equation}
\label{eq:cut-off-delta}
    \Lambda^{\delta+2}\approx (36\pi)^3\left(\frac{ 2^\delta\, \delta}{S_{\delta-1}}\right)\,\left(\frac{f_4^2 \mu_1^\delta}{\psi_1^{2\delta}}\right)\,.
\end{equation}

\bibliographystyle{BiblioStyle}
\bibliography{DraftBiblio}

\providecommand{\href}[2]{#2}\begingroup\raggedright\begin{thebibliography}{10}

\bibitem{WITTEN1984351}
E.~Witten, {\it Some properties of o(32) superstrings},  Physics Letters B {\bf 149} (1984), no.~4 351--356.

\bibitem{Arvanitaki:2009fg}
A.~Arvanitaki, S.~Dimopoulos, S.~Dubovsky, N.~Kaloper, and J.~March-Russell, {\it {String Axiverse}},  Phys. Rev. D {\bf 81} (2010) 123530, [\href{http://arxiv.org/abs/0905.4720}{{\tt arXiv:0905.4720}}].

\bibitem{Heidenreich:2020pkc}
B.~Heidenreich, J.~McNamara, M.~Montero, M.~Reece, T.~Rudelius, and I.~Valenzuela, {\it {Chern-Weil global symmetries and how quantum gravity avoids them}},  JHEP {\bf 11} (2021) 053, [\href{http://arxiv.org/abs/2012.00009}{{\tt arXiv:2012.00009}}].

\bibitem{AxionLimits}
C.~O'Hare, ``cajohare/axionlimits: Axionlimits.'' \url{https://cajohare.github.io/AxionLimits/}, July, 2020.

\bibitem{Gavela:2023tzu}
B.~Gavela, P.~Qu\'\i{}lez, and M.~Ramos, {\it {The QCD axion sum rule}},  \href{http://arxiv.org/abs/2305.15465}{{\tt arXiv:2305.15465}}.

\bibitem{Kaluza:1921tu}
T.~Kaluza, {\it {Zum Unit\"atsproblem der Physik}},  Sitzungsber. Preuss. Akad. Wiss. Berlin (Math. Phys. ) {\bf 1921} (1921) 966--972, [\href{http://arxiv.org/abs/1803.08616}{{\tt arXiv:1803.08616}}].

\bibitem{Klein:1926tv}
O.~Klein, {\it {Quantum Theory and Five-Dimensional Theory of Relativity. (In German and English)}},  Z. Phys. {\bf 37} (1926) 895--906.

\bibitem{Dienes:1999gw}
K.~R. Dienes, E.~Dudas, and T.~Gherghetta, {\it {Invisible axions and large radius compactifications}},  Phys. Rev. D {\bf 62} (2000) 105023, [\href{http://arxiv.org/abs/hep-ph/9912455}{{\tt hep-ph/9912455}}].

\bibitem{DiLella:2000dn}
L.~Di~Lella, A.~Pilaftsis, G.~Raffelt, and K.~Zioutas, {\it {Search for solar Kaluza-Klein axions in theories of low scale quantum gravity}},  Phys. Rev. D {\bf 62} (2000) 125011, [\href{http://arxiv.org/abs/hep-ph/0006327}{{\tt hep-ph/0006327}}].

\bibitem{Flacke:2006ad}
T.~Flacke, B.~Gripaios, J.~March-Russell, and D.~Maybury, {\it {Warped axions}},  JHEP {\bf 01} (2007) 061, [\href{http://arxiv.org/abs/hep-ph/0611278}{{\tt hep-ph/0611278}}].

\bibitem{Anastasopoulos:2018uyu}
P.~Anastasopoulos, P.~Betzios, M.~Bianchi, D.~Consoli, and E.~Kiritsis, {\it {Emergent/Composite axions}},  JHEP {\bf 10} (2019) 113, [\href{http://arxiv.org/abs/1811.05940}{{\tt arXiv:1811.05940}}].

\bibitem{Cox:2019rro}
P.~Cox, T.~Gherghetta, and M.~D. Nguyen, {\it {A Holographic Perspective on the Axion Quality Problem}},  JHEP {\bf 01} (2020) 188, [\href{http://arxiv.org/abs/1911.09385}{{\tt arXiv:1911.09385}}].

\bibitem{Gherghetta:2020keg}
T.~Gherghetta, V.~V. Khoze, A.~Pomarol, and Y.~Shirman, {\it {The Axion Mass from 5D Small Instantons}},  JHEP {\bf 03} (2020) 063, [\href{http://arxiv.org/abs/2001.05610}{{\tt arXiv:2001.05610}}].

\bibitem{Bonnefoy:2020llz}
Q.~Bonnefoy, P.~Cox, E.~Dudas, T.~Gherghetta, and M.~D. Nguyen, {\it {Flavoured Warped Axion}},  JHEP {\bf 04} (2021) 084, [\href{http://arxiv.org/abs/2012.09728}{{\tt arXiv:2012.09728}}].

\bibitem{Gendler:2024gdo}
N.~Gendler and C.~Vafa, {\it {Axions in the Dark Dimension}},  \href{http://arxiv.org/abs/2404.15414}{{\tt arXiv:2404.15414}}.

\bibitem{Agrawal:2024ejr}
P.~Agrawal, M.~Nee, and M.~Reig, {\it {Axion Couplings in Heterotic String Theory}},  \href{http://arxiv.org/abs/2410.03820}{{\tt arXiv:2410.03820}}.

\bibitem{Craig:2024dnl}
N.~Craig and M.~Kongsore, {\it {High-Quality Axions from Higher-Form Symmetries in Extra Dimensions}},  \href{http://arxiv.org/abs/2408.10295}{{\tt arXiv:2408.10295}}.

\bibitem{Dienes:2011ja}
K.~R. Dienes and B.~Thomas, {\it {Dynamical Dark Matter: I. Theoretical Overview}},  Phys. Rev. D {\bf 85} (2012) 083523, [\href{http://arxiv.org/abs/1106.4546}{{\tt arXiv:1106.4546}}].

\bibitem{Dienes:2011sa}
K.~R. Dienes and B.~Thomas, {\it {Dynamical Dark Matter: II. An Explicit Model}},  Phys. Rev. D {\bf 85} (2012) 083524, [\href{http://arxiv.org/abs/1107.0721}{{\tt arXiv:1107.0721}}].

\bibitem{Dienes:2012jb}
K.~R. Dienes and B.~Thomas, {\it {Phenomenological Constraints on Axion Models of Dynamical Dark Matter}},  Phys. Rev. D {\bf 86} (2012) 055013, [\href{http://arxiv.org/abs/1203.1923}{{\tt arXiv:1203.1923}}].

\bibitem{Randall:1999ee}
L.~Randall and R.~Sundrum, {\it {A Large mass hierarchy from a small extra dimension}},  Phys. Rev. Lett. {\bf 83} (1999) 3370--3373, [\href{http://arxiv.org/abs/hep-ph/9905221}{{\tt hep-ph/9905221}}].

\bibitem{Randall:1999vf}
L.~Randall and R.~Sundrum, {\it {An Alternative to compactification}},  Phys. Rev. Lett. {\bf 83} (1999) 4690--4693, [\href{http://arxiv.org/abs/hep-th/9906064}{{\tt hep-th/9906064}}].

\bibitem{Gendler:2023kjt}
N.~Gendler, D.~J.~E. Marsh, L.~McAllister, and J.~Moritz, {\it {Glimmers from the axiverse}},  JCAP {\bf 09} (2024) 071, [\href{http://arxiv.org/abs/2309.13145}{{\tt arXiv:2309.13145}}].

\bibitem{Agrawal:2022lsp}
P.~Agrawal, M.~Nee, and M.~Reig, {\it {Axion couplings in grand unified theories}},  JHEP {\bf 10} (2022) 141, [\href{http://arxiv.org/abs/2206.07053}{{\tt arXiv:2206.07053}}].

\bibitem{Grebenkov_2020}
D.~S. Grebenkov, {\it A physicist's guide to explicit summation formulas involving zeros of bessel functions and related spectral sums},  Reviews in Mathematical Physics {\bf 33} (nov, 2020) 2130002.

\bibitem{Flacke:2006re}
T.~Flacke and D.~Maybury, {\it {Aspects of Axion Phenomenology in a slice of AdS(5)}},  JHEP {\bf 03} (2007) 007, [\href{http://arxiv.org/abs/hep-ph/0612126}{{\tt hep-ph/0612126}}].

\bibitem{Buyukdag:2019lhh}
Y.~Buyukdag, K.~R. Dienes, T.~Gherghetta, and B.~Thomas, {\it {Partially Composite Dynamical Dark Matter}},  Phys. Rev. D {\bf 101} (2020), no.~7 075054, [\href{http://arxiv.org/abs/1912.10588}{{\tt arXiv:1912.10588}}].

\bibitem{deGiorgi:2021xvm}
A.~de~Giorgi and S.~Vogl, {\it {Dark matter interacting via a massive spin-2 mediator in warped extra-dimensions}},  JHEP {\bf 11} (2021) 036, [\href{http://arxiv.org/abs/2105.06794}{{\tt arXiv:2105.06794}}].

\bibitem{Callin:2004py}
P.~Callin and F.~Ravndal, {\it {Higher order corrections to the Newtonian potential in the Randall-Sundrum model}},  Phys. Rev. D {\bf 70} (2004) 104009, [\href{http://arxiv.org/abs/hep-ph/0403302}{{\tt hep-ph/0403302}}].

\bibitem{Hoskins:1985tn}
J.~K. Hoskins, R.~D. Newman, R.~Spero, and J.~Schultz, {\it {Experimental tests of the gravitational inverse square law for mass separations from 2-cm to 105-cm}},  Phys. Rev. D {\bf 32} (1985) 3084--3095.

\bibitem{Bordag:2001qi}
M.~Bordag, U.~Mohideen, and V.~M. Mostepanenko, {\it {New developments in the Casimir effect}},  Phys. Rept. {\bf 353} (2001) 1--205, [\href{http://arxiv.org/abs/quant-ph/0106045}{{\tt quant-ph/0106045}}].

\bibitem{Mostepanenko:2001fx}
V.~M. Mostepanenko and M.~Novello, {\it {Constraints on nonNewtonian gravity from the Casimir force measurements between two crossed cylinders}},  Phys. Rev. D {\bf 63} (2001) 115003, [\href{http://arxiv.org/abs/hep-ph/0101306}{{\tt hep-ph/0101306}}].

\bibitem{Chiaverini:2002cb}
J.~Chiaverini, S.~J. Smullin, A.~A. Geraci, D.~M. Weld, and A.~Kapitulnik, {\it {New experimental constraints on nonNewtonian forces below 100 microns}},  Phys. Rev. Lett. {\bf 90} (2003) 151101, [\href{http://arxiv.org/abs/hep-ph/0209325}{{\tt hep-ph/0209325}}].

\bibitem{Long:2003dx}
J.~C. Long, H.~W. Chan, A.~B. Churnside, E.~A. Gulbis, M.~C.~M. Varney, and J.~C. Price, {\it {Upper limits to submillimeter-range forces from extra space-time dimensions}},  Nature {\bf 421} (2003) 922--925, [\href{http://arxiv.org/abs/hep-ph/0210004}{{\tt hep-ph/0210004}}].

\bibitem{Chen:2014oda}
Y.~J. Chen, W.~K. Tham, D.~E. Krause, D.~Lopez, E.~Fischbach, and R.~S. Decca, {\it {Stronger Limits on Hypothetical Yukawa Interactions in the 30\textendash{}8000 nm Range}},  Phys. Rev. Lett. {\bf 116} (2016), no.~22 221102, [\href{http://arxiv.org/abs/1410.7267}{{\tt arXiv:1410.7267}}].

\bibitem{Tan:2016vwu}
W.-H. Tan, S.-Q. Yang, C.-G. Shao, J.~Li, A.-B. Du, B.-F. Zhan, Q.-L. Wang, P.-S. Luo, L.-C. Tu, and J.~Luo, {\it {New Test of the Gravitational Inverse-Square Law at the Submillimeter Range with Dual Modulation and Compensation}},  Phys. Rev. Lett. {\bf 116} (2016), no.~13 131101.

\bibitem{Lee:2020zjt}
J.~G. Lee, E.~G. Adelberger, T.~S. Cook, S.~M. Fleischer, and B.~R. Heckel, {\it {New Test of the Gravitational $1/r^2$ Law at Separations down to 52 $\mu$m}},  Phys. Rev. Lett. {\bf 124} (2020), no.~10 101101, [\href{http://arxiv.org/abs/2002.11761}{{\tt arXiv:2002.11761}}].

\bibitem{Hannestad:2003yd}
S.~Hannestad and G.~G. Raffelt, {\it {Supernova and neutron star limits on large extra dimensions reexamined}},  Phys. Rev. D {\bf 67} (2003) 125008, [\href{http://arxiv.org/abs/hep-ph/0304029}{{\tt hep-ph/0304029}}]. [Erratum: Phys.Rev.D 69, 029901 (2004)].

\bibitem{Cembranos:2017vgi}
J.~A.~R. Cembranos, A.~L. Maroto, and H.~Villarrubia-Rojo, {\it {Constraints on hidden gravitons from fifth-force experiments and stellar energy loss}},  JHEP {\bf 09} (2017) 104, [\href{http://arxiv.org/abs/1706.07818}{{\tt arXiv:1706.07818}}].

\bibitem{Jodlowski:2023yne}
K.~Jod\l{}owski, {\it {Probing some photon portals to new physics at intensity frontier experiments}},  Phys. Rev. D {\bf 108} (2023), no.~11 115017, [\href{http://arxiv.org/abs/2305.05710}{{\tt arXiv:2305.05710}}].

\bibitem{dEnterria:2023npy}
D.~d'Enterria, M.~A. Tamlihat, L.~Schoeffel, H.-S. Shao, and Y.~Tayalati, {\it {Collider constraints on massive gravitons coupling to photons}},  Phys. Lett. B {\bf 846} (2023) 138237, [\href{http://arxiv.org/abs/2306.15558}{{\tt arXiv:2306.15558}}].

\bibitem{DiLuzio:2017ogq}
L.~Di~Luzio, F.~Mescia, E.~Nardi, P.~Panci, and R.~Ziegler, {\it {Astrophobic Axions}},  Phys. Rev. Lett. {\bf 120} (2018), no.~26 261803, [\href{http://arxiv.org/abs/1712.04940}{{\tt arXiv:1712.04940}}].

\bibitem{Giudice:2016yja}
G.~F. Giudice and M.~McCullough, {\it {A Clockwork Theory}},  JHEP {\bf 02} (2017) 036, [\href{http://arxiv.org/abs/1610.07962}{{\tt arXiv:1610.07962}}].

\bibitem{polytopes}
M.~Brion and M.~Vergne, {\it Lattice points in simple polytopes},  Journal of the American Mathematical Society {\bf 10} (1997), no.~2 371--392.

\bibitem{Goldberger:1999wh}
W.~D. Goldberger and M.~B. Wise, {\it {Bulk fields in the Randall-Sundrum compactification scenario}},  Phys. Rev. D {\bf 60} (1999) 107505, [\href{http://arxiv.org/abs/hep-ph/9907218}{{\tt hep-ph/9907218}}].

\bibitem{Horvat:2003xd}
R.~Horvat, M.~Krcmar, and B.~Lakic, {\it {CERN Axion Solar Telescope as a probe of large extra dimensions}},  Phys. Rev. D {\bf 69} (2004) 125011, [\href{http://arxiv.org/abs/astro-ph/0312030}{{\tt astro-ph/0312030}}].

\bibitem{Bastero-Gil:2021oky}
M.~Bastero-Gil, C.~Beaufort, and D.~Santos, {\it {Solar axions in large extra dimensions}},  JCAP {\bf 10} (2021) 048, [\href{http://arxiv.org/abs/2107.13337}{{\tt arXiv:2107.13337}}].

\bibitem{NEWS-G:2021vfh}
{\bf NEWS-G} Collaboration, Q.~Arnaud {\em et.~al.}, {\it {Solar Kaluza-Klein axion search with NEWS-G}},  Phys. Rev. D {\bf 105} (2022), no.~1 012002, [\href{http://arxiv.org/abs/2109.03562}{{\tt arXiv:2109.03562}}].

\bibitem{Bedi:2024wqg}
R.~Bedi, T.~Gherghetta, C.~Grojean, G.~Guedes, J.~Kley, and P.~N.~H. Vuong, {\it {Small instanton-induced flavor invariants and the axion potential}},  JHEP {\bf 06} (2024) 156, [\href{http://arxiv.org/abs/2402.09361}{{\tt arXiv:2402.09361}}].

\bibitem{Chanowitz:1978uj}
M.~S. Chanowitz, M.~A. Furman, and I.~Hinchliffe, {\it {Weak Interactions of Ultraheavy Fermions}},  Phys. Lett. B {\bf 78} (1978) 285.

\bibitem{ATLAS:2023tgo}
{\bf ATLAS} Collaboration, G.~Aad {\em et.~al.}, {\it {Determination of the strong coupling constant from transverse energy$-$energy correlations in multijet events at $\sqrt{s} = 13$ TeV with the ATLAS detector}},  JHEP {\bf 07} (2023) 085, [\href{http://arxiv.org/abs/2301.09351}{{\tt arXiv:2301.09351}}].

\end{thebibliography}\endgroup

\end{document}